\def\draft{1}
\documentclass[11pt]{article}
\usepackage{macros}

\title{A Quasipolynomial Quantum Algorithm\\ for the Vector Dihedral Coset Problem\\
with Smooth Modulus}
\author{Aparna Gupte\\MIT}
\date{\today}

\begin{document}

\maketitle

\begin{abstract}
We give a quantum algorithm for the vector dihedral coset problem over $\ZZ_q^n$, in which the goal is to recover $\vecs \in \ZZ_q^n$ given independent samples of the form $\frac{1}{\sqrt{2}}\bigl(\ket{0,\vecx}+\ket{1,\vecx+\vecs}\bigr)$, for $\vecx \sim \ZZ_q^n$. When the modulus $q$ is $B$-smooth, the algorithm uses $n^{O(B^{1+o(1)}\log q)}B^{O(B^{2+o(1)}\log q)}$ samples and time. For $B = \polylog(n)$ and $q=\poly(n)$, this gives a quasipolynomial time- and sample-complexity. This extends the previously known quasipolynomial-time result for power-of-two moduli (Bonnetain--Naya-Plasencia, ASIACRYPT 2018; Bai--Jangir--Kirshanova--Ngo--Youmans, CRYPTO 2025).

The algorithm proceeds by reducing the modulus one prime factor $p | q$ at a time. The key ingredient is a sieving procedure that clears the non-linear terms in the phase modulo $p$ via an algorithm of Ivanyos and Santha (TCS 2017) that finds non-zero $\vecu \in \FF_p^\ell$ such that $\sum_{i \in [\ell]} u_i^d \veca_i = 0 \mod p$ for degrees $1 \le d \le p-1$, given vectors $\veca_1, \ldots, \veca_\ell \in \FF_p^n$ as input.

The algorithm also applies to the extrapolated dihedral coset problem (EDCP) introduced by Brakerski--Kirshanova--Stehlé--Wen (PKC 2018), which is known to be equivalent to LWE in the polynomial sample regime. However, our result does not yield a new attack on LWE in any regime, precisely because the known reduction from LWE to EDCP provides only polynomially many clean samples, whereas our algorithm requires quasipolynomially many samples.

The main algorithm was obtained in a GPT-5.6 Sol conversation prompted by the author; the author is responsible for all statements, proofs, and correctness. Concurrent work of Boucher--Fouque--Shen gives an algorithm with similar guarantees for prime-power moduli.
\end{abstract}

\newpage
\tableofcontents
\newpage

\section{Introduction}
The \emph{dihedral coset problem} $\DCP_q^m$ asks to recover a secret $s\in\ZZ_q$ from $m$ independent quantum samples of the form
\begin{align*}
    \frac{1}{\sqrt2}
    \left(
        \ket{0}\ket{x_i}
        +
        \ket{1}\ket{x_i+s}
    \right),
    \qquad
    x_i\gets\ZZ_q,\ i\in[m],
\end{align*}
where the offsets $x_i$ are uniformly random and unknown to the solver. States of this form naturally appear in the ``standard approach'' for the dihedral hidden subgroup problem and for the hidden shift problem over $\ZZ_q$~\cite{Kuperberg05,Regev04DHSP}; solving DCP in polynomial time would imply polynomial-time quantum algorithms for both these problems. The best algorithm we know for all three of these problems is due to Kuperberg~\cite{Kuperberg05}, and has subexponential time and sample complexity $2^{O(\sqrt{\log q})}$.

Interestingly, DCP upper-bounds the hardness of the computational problems underlying several of our candidate post-quantum cryptographic assumptions. 
Key recovery for a cryptographic abelian group action is an instance of the hidden shift problem~\cite{ChildsJaoSoukharev14}, and Kuperberg-style DCP algorithms give the best known attacks for group action-based cryptography~\cite{BonnetainSchrottenloher20,Peikert20}.
Similarly, DCP is closely related to Learning With Errors (LWE), a central problem at the heart of lattice-based cryptography.
Regev~\cite{Regev04} showed that $\poly(n)$-$\mathsf{uSVP}$ (and therefore also LWE) is no harder than DCP, by giving a polynomial-time quantum reduction to $\DCP_q^m$. Crucially, this reduction produces only polynomially many noiseless DCP samples, rendering Kuperberg's subexponential algorithm irrelevant.
Information-theoretically, however, $m=O(\log q)$ samples already determine $s$~\cite{EttingerHoyerKnill04}, and an algorithm with polynomial time- and sample-complexity for DCP would give a polynomial-time attack on lattice-based cryptography. The reduction, however, runs in one direction only: LWE might potentially be a lot easier than DCP.

To close this potential gap in hardness, Brakerski, Kirshanova, Stehl\'e and Wen~\cite{brakerski2018learning} introduced the \emph{extrapolated} dihedral coset problem $\EDCP_{n,q,f}^m$. It is parameterized additionally by a dimension $n\in\NN$ and an amplitude function $f:\ZZ\to\CC$ with $\sum_{k\in\ZZ}\abs{f(k)}^2=1$, and asks to recover a secret \emph{vector} $\vecs\in\ZZ_q^n$ from $m$ independent quantum states of the form
\begin{align*}
    \sum_{k \in \ZZ} f(k) \ket{k} \ket{\vecx_i + k \cdot \vecs}, \qquad
    \vecx_i\in\ZZ_q^n,\ i\in[m].
\end{align*}
Taking $n=1$ and $f(k) = \frac{1}{\sqrt{2}} \cdot \mathbbm{1}[k \in \{0,1\}]$ gives two-point DCP states. Writing $\mathsf{U}\text{-}\EDCP_{n,q,r}^m$ for the
\emph{uniform} case, in which $f$ is the normalized indicator of
$\{0,1,\ldots,r-1\}$, Brakerski et al.\ showed that
$\mathsf{U}\text{-}\EDCP_{n,q,r}^m$ and $\mathsf{LWE}_{n,q,\alpha}$ are
equivalent up to polynomial losses in parameters under polynomial-time quantum reductions when
$m=\poly(n\log q)$ and $r=\poly(n\log q)/\alpha$. 

Throughout this paper, we specialize to the two-point amplitude $f(k)=\frac{1}{\sqrt{2}}\mathbbm{1}[k\in\{0,1\}]$, and call this problem \emph{vector dihedral coset problem}, denoted as $\DCP_{n,q}^m$. This problem is known to be as hard as $\mathsf{U}\text{-}\EDCP_{n,q,r}^m$~\cite[Lemma 14]{BaiEtAl25}.
Our goal is thus to recover $\vecs$ from independent samples of the form
\begin{align*}
    \frac{1}{\sqrt2}
    \left(
        \ket{0}\ket{\vecx_i}
        +
        \ket{1}\ket{\vecx_i+\vecs}
    \right),
    \qquad
    \vecx _i \gets\ZZ_q^n, i \in [m].
\end{align*}

For general moduli, the best known algorithm is still Kuperberg's, which solves $\DCP_q^m$ in time and samples $2^{O(\sqrt{\log q})}$ \cite{Kuperberg05}; applied to the group $\ZZ_q^n$ it solves $\DCP_{n,q}^m$ in time $2^{O(\sqrt{n\log q})}$. For \emph{structured} moduli, however, one can do considerably better.  When $q=p$ is a fixed prime, Friedl, Ivanyos, Magniez, Santha and Sen~\cite{FriedlEtAl14} solve the problem in polynomial time, and when $q=2^t$ is a power of two, Bai, Jangir, Kirshanova, Ngo and Youmans~\cite{BaiEtAl25} give a quasipolynomial-time algorithm for $q=\poly(n)$. Both exploit special structure of the modulus, which raises the question:
\begin{center}
\emph{For which moduli $q$ does vector DCP admit an algorithm that beats Kuperberg's?}
\end{center}

\subsection{Our main result and idea}
Call a modulus \emph{$B$-smooth} if all of its prime factors are at most $B$. Our main result is an algorithm for $\DCP_{n,q}^m$ whose cost is governed by the largest prime factor of $q$.

\begin{informaltheorem}[\Cref{thm:edcp-correctness}]
Let $q>1$ be $B$-smooth.  There is a quantum algorithm that recovers
$\vecs\in\ZZ_q^n$ from two-point vector DCP states with success probability
at least $1-\varepsilon$, using
\begin{align*}
    n^{O(B^{1+o(1)}\log q)}B^{O(B^{2+o(1)}\log q)}
    \poly\!\left(\log\frac{2}{\varepsilon}\right)
\end{align*}
samples and time. This is quasipolynomial when $B=\polylog(n)$ and $q=\poly(n)$.
\end{informaltheorem}

The work of Boucher, Fouque and Shen~\cite{boucher2026cyclotomic} uses different techniques to obtain an algorithm for prime-power moduli $q = p^t$. We handle smooth moduli with possibly different prime factors. We compare our results in more detail in \Cref{sec:related-work}.

Our algorithm proceeds by reducing the modulus one prime factor $p | q$ at a time. The main idea is that this can be done by a sieving procedure that clears the non-linear terms in the phase modulo $p$, one degree at a time, via a subroutine that finds non-zero $\vecu \in \FF_p^\ell$ such that $\sum_{i \in [\ell]} u_i^d \veca_i = 0 \mod p$, given vectors $\veca_1, \ldots, \veca_\ell \in \FF_p^n$. We use the diagonal-equation algorithm of Ivanyos and Santha~\cite{IvanyosSantha16} for this purpose; the subset sum algorithms of Imran and Ivanyos~\cite{ImranIvanyos23} or Kothari--O'Donnell--Wu~\cite{KothariODonnellWu26} give alternative ways to find these relations with slightly worse dependence on $B$ in the exponent of $n$. We describe our techniques in more detail in~\Cref{sec:tech-overview}.

We describe our main algorithm (\Cref{algo:main}) and state the main result (\Cref{thm:finite-abelian-hidden-shift}) in terms of DCP for the generalized dihedral group $\dih(G)$, where $\ZZ_q^n$ is replaced with a general finite abelian group $G$ to which we have black-box access. This allows us to highlight the structure of $\ZZ_q^n$ that enables a faster algorithm (see \Cref{cor:abelian-smooth-exponent}).

\subsection{Related work}\label{sec:related-work}
The algorithm of Kuperberg~\cite{Kuperberg05} gives a subexponential-time quantum algorithm for the \emph{hidden shift problem} over arbitrary finite abelian groups: given oracle access to injective functions $f_0,f_1:G\to X$ satisfying $f_0(x)=f_1(x+s)$ for some unknown $s\in G$, the goal is to recover $s$. Querying these functions in superposition and measuring the output register yields states of the form $\frac{1}{\sqrt 2}(\ket{0,x}+\ket{1,x+s})$, where $x\in G$ is uniformly random. These are coset states of the generalized dihedral group $\dih(G)$, defined in \Cref{def:generalized-dihedral}.

Specializing to $G=\ZZ_q^n$ yields the vector dihedral coset problem, which Kuperberg's algorithm solves in time and sample complexity $2^{O(\sqrt{n\log q})}$, for arbitrary moduli $q$. In this setting, Kuperberg's algorithm works by first performing a quantum Fourier transform to convert DCP samples into Fourier samples of the form $\left(\veca, \frac{1}{\sqrt{2}} (\ket{0} + \omega_q^{\inner{\veca, \vecs}} \ket{1})\right)$, and then performs a ``sieving'' step that iteratively combines these Fourier samples until $\veca \in (q/2) \ZZ_q^n$. At this point, the least significant bits of $\vecs$ can be learned by a Hadamard transform and Gaussian elimination.

\paragraph{Memory-efficient algorithms.}
In addition to requiring subexponential samples and runtime, Kuperberg's original algorithm requires $2^{O(\sqrt{\log |G|})}$ quantum memory. Regev reduced the space complexity to $\poly(\log |G|)$ for cyclic $2$-groups, at the expense of increasing the time and sample complexity to $2^{O(\sqrt{\log |G|\log\log |G|})}$~\cite{Regev04DHSP}. Kuperberg's later collimation sieve recovered the original $2^{O(\sqrt{\log |G|})}$ time bound while requiring only $O(\log |G|)$ qubits, at the cost of subexponential $2^{O(\sqrt{\log |G|})}$ classical memory accessible in quantum superposition~\cite{Kuperberg13}. Peikert~\cite{Peikert20} extended the collimation sieve to arbitrary cyclic groups and analyzed its concrete costs for CSIDH, an isogeny-based key-exchange protocol built from a commutative group action on supersingular elliptic curves.

\paragraph{Time- and sample-efficient algorithms for structured groups.}
While the preceding works reduce the space complexity of Kuperberg's algorithm, additional group structure can be exploited to get algorithms that run faster and use fewer samples. The simplest case is $G=\ZZ_2^n$, where hidden shift reduces to Simon's problem and can be solved in $\poly(n)$ time using $\poly(n)$ samples by Fourier sampling and then Gaussian elimination~\cite{simon1997power}. Friedl, Ivanyos, Magniez, Santha, and Sen~\cite{FriedlEtAl14} generalized this result to $G=\FF_p^n$ for any prime $p$, achieving time complexity $n^{O(p)}p^{O(p)}$. Their key observation is that quantum Fourier sampling can be used to obtain random linear \emph{disequations} of the form $\langle a,s\rangle\neq 0\bmod p$. By Fermat's little theorem, these can then be converted into linear equations in the degree-$(p-1)$ monomials of $s$, allowing the secret to be recovered using linear algebra~\cite{Ivanyos08}.

Friedl et al.~\cite{FriedlEtAl14} also extend their algorithm to more general groups using a recursion-friendly generalization of the hidden shift problem, in which the secret shift can be an entire coset, rather than just a single element.\footnote{Friedl et al.~call this generalization the Translating Coset problem.} Given hidden shift oracles $f_0,f_1:G\to X$, they first prepare the states
\[
\ket{f_b}:=\frac{1}{\sqrt{|G|}}\sum_{x\in G}\ket{x,f_b(x)},
\]
so that the hidden shift relation becomes $s\cdot\ket{f_0}=\ket{f_1}$, where $s$ acts by translating the first register by $s$; the recursive iterations of the algorithm prepare states with a coset of hidden shifts. Their recursive algorithm reduces this quantum-state version of hidden shift to instances over subgroups $H\le G$ such that the quotient $G/H \cong \FF_p^r$ is amenable to their $\FF_p^n$ algorithm.
In particular, for $G=\ZZ_q^n$ with all prime factors of $q$ bounded by $B$, their algorithm runs in time $(n\log q)^{O_B(\log q)}$, which is quasipolynomial for constant $B$ and $q=\poly(n)$.\footnote{They study hidden shift over nonabelian groups as well, and refer to this variant as the \emph{hidden translation} problem. They obtain polynomial-time algorithms for smoothly solvable groups, quasipolynomial-time algorithms for solvable groups of constant exponent, and subexponential-time algorithms for arbitrary solvable groups. While it is not important for this work, a group $G$ is \emph{solvable} if it admits a subnormal series $G = G_0 \ge G_1 \ge \ldots \ge \{0\}$  with abelian quotients $G_i/G_{i+1}$, and its \emph{exponent} is the least common multiple of the orders of its elements.} However, their recursion uses copies of the full function states $\ket{f_0},\ket{f_1}$ and coherent operations on them, rather than just independent dihedral coset states. It therefore does not directly yield an algorithm for the vector DCP problem considered here.\footnote{We remark that this distinction does not rule out cryptographic applications: group-action-based cryptosystems provide coherent access to the hidden shift functions, and adaptations of Regev's reduction may yield noisy versions of the function states required by Friedl et al. Nevertheless, this recursive algorithm in Friedl et al.'s work appears to have received relatively little attention in the cryptographic literature, possibly because it is formulated in terms of Translating Coset rather than DCP.}

Our algorithm can be viewed as a counterpart to the hidden shift algorithm of Friedl et al.~\cite{FriedlEtAl14} that works directly with independent DCP samples, rather than coherent access to hidden shift function oracles. The two approaches also use complementary classical subroutines: Friedl et al.\ solve linear disequations over $\FF_p^n$, whereas we solve diagonal equations over $\FF_p^n$, with subset sum as a special case. This suggests a duality between their approach and ours. The duality of the two problems can also be seen in the work of Chen, Liu, and Zhandry~\cite{ChenLiuZhandry22}, which reduces subset sum over $\FF_p^n$ to solving disequations via Fourier transform and unambiguous state discrimination.

\paragraph{Algorithms for DCP over power-of-two moduli.}
For $q=2^k$, faster algorithms are known that work directly with independent DCP states, achieving time and sample complexity $n^{O(\log q)}$. Bonnetain and Naya-Plasencia~\cite{BonnetainNayaPlasencia18} introduced the \emph{Simon-meets-Kuperberg} algorithm, which interpolates between Simon's polynomial-time algorithm when $k=1$ and Kuperberg's subexponential-time algorithm when $n=1$. Cs\'aji~\cite{Csaji21} subsequently gave a rigorous, space-efficient algorithm achieving $n^{O(\log q)}$ time and sample complexity. More recently, Bai, Jangir, Kirshanova, Ngo, and Youmans~\cite{BaiEtAl25} gave a provable variant of Simon-meets-Kuperberg that also handles extrapolated dihedral coset states (EDCP), achieving the same asymptotic complexity for power-of-two moduli. These algorithms exploit linear dependencies over $\FF_2$ to progressively reduce the modulus by factors of two.

\paragraph{Beyond powers of two.}
Castryck, Dooms, Emerencia, and Lemmens~\cite{CastryckEtAl21} combined the power-of-two sieving techniques of \cite{BonnetainNayaPlasencia18} with the $\FF_p^n$ algorithm of Friedl et al.~\cite{FriedlEtAl14}, to give a $n^{O_p(\log q)}$ algorithm for vector DCP over moduli of the form $q = p \cdot 2^k$. They also extend Peikert's collimation sieve to arbitrary finite abelian groups. Combining these techniques, they obtain a hidden shift algorithm for any finite abelian group $G$ running in time $\poly(\log |G|)\cdot 2^{O(\sqrt{\log |2^k pG|})}$, using only polynomially many qubits and quantum-accessible classical memory.

\paragraph{Concurrent work of Boucher, Fouque, and Shen~\cite{boucher2026cyclotomic}.}
The recent concurrent work by Boucher, Fouque, and Shen~\cite{boucher2026cyclotomic} extends the Simon-meets-Kuperberg framework to vector DCP over prime-power moduli $q=p^k$, achieving $n^{O_p(\log q)}$ time and sample complexity. 
They do this by (1) defining the \emph{cyclotomic coset problem} (CCP), a $p$-ary generalization of the binary \emph{dihedral} coset problem (DCP), (2) generalizing the Simon-meets-Kuperberg framework to CCP, and (3) giving a reduction from DCP to CCP.

The dihedral group $\dih(\ZZ_q^n)$ can be viewed as a semidirect product $\ZZ_q^n \rtimes \ZZ_2$, with group law $ (\vecx, a) \times (\vecy, b) = (\vecx + (-1)^a \vecy, a + b),$ and DCP asks to recover the subgroup $\{(0, 0), (\vecs,1)\}$ from the coset states $\frac{1}{\sqrt2}\left(\ket{0}\ket{\vecx} + \ket{1}\ket{\vecx+\vecs} \right)$.
Boucher, Fouque, and Shen~\cite{boucher2026cyclotomic} define CCP analogously for the cyclotomic group $\ZZ_q[\zeta_p]^n \rtimes \ZZ_p$, with group law
\begin{align*}
    (\vecx, a) \times (\vecy, b) = (\vecx + \zeta_p^a \vecy, a + b),
\end{align*}
where $\zeta_p$ is a primitive $p$th root of unity. They define CCP to be the task of recovering the subgroup $\left\{\left( \lambda_a \cdot \vecs, a\right) : a \in \ZZ_p\right\}$, where $\lambda_a := \sum_{j =0}^{a-1} \zeta_p^j $, from coset states of the form
\begin{align*}
    \frac{1}{\sqrt p} \sum_{a \in \ZZ_p} \ket{a} \ket{\vecx + \lambda_a \cdot \vecs}.
\end{align*}
They then extend the Simon-meets-Kuperberg framework to work for CCP states, via a subset sum algorithm over $\FF_p^n$ of \cite{ImranIvanyos23}.

Our algorithm also combines Fourier samples using relations among their labels. We work directly with the Fourier samples obtained from the DCP coset states, clearing nonlinear phase terms modulo $p$ one degree at a time to get ``semi-clean'' states whose phase terms are linear modulo $p$. Such ``semi-clean'' Fourier samples can be used to reduce the modulus by a factor of $p$; this process can then be repeated.
In the language of our algorithm, the cyclotomic states of \cite{boucher2026cyclotomic} are already semi-clean modulo $\zeta_p-1$: their degree-$d$ binomial coefficient is $(\zeta_p-1)^{d-1}y$, so all nonlinear coefficients vanish modulo $\zeta_p-1$, whereas our algorithm achieves this linearity by explicitly cancelling the nonlinear coefficients of arbitrary phase-label maps.

Although their current analysis is restricted to prime power moduli, we suspect that it might be possible to extend the sieve of \cite{boucher2026cyclotomic} to smooth moduli, perhaps by giving a reduction that prepares coset states for one cyclotomic group $\ZZ_q[\zeta_{p_1}]^n \rtimes \FF_{p_1}$ from coset states for another cyclotomic group $\ZZ_q[\zeta_{p_2}]^n \rtimes \FF_{p_2}$, where $p_1, p_2 | q$, but $p_1 \neq p_2$; however this remains to be checked. Our construction directly works with $p$-ary Fourier samples, and handles the more general smooth modulus setting.

While \cite{boucher2026cyclotomic} uses the subset sum algorithm of \cite{ImranIvanyos23}, we observe that by instead using the algorithm of Ivanyos and Santha~\cite{IvanyosSantha16} to solve the more general degree-$d$ problem of finding nonzero $\vecu$ such that $\sum_i u_i^d\veca_i=\veczero$, we get a slightly better dependence on $B$ in the exponent of $n$ in the sample and time complexity. For $q=p^t$, our time and sample bound from~\Cref{thm:edcp-correctness} is
\[
    n^{O(p^{1+o(1)}\log q)}p^{O(p^{2+o(1)}\log q)}.
\]
The theorem of \cite{boucher2026cyclotomic} states a $2^{O_p(\log n\log q)}$ bound for fixed $p$, but to the best of our understanding, making the dependence on $p$ explicit gives a $n^{O(p^2\log q)}p^{O(p^2\log p\,\log q)}$ algorithm.

\paragraph{Subset sum and diagonal polynomial equations over $\FF_p^n$.} Our algorithm needs a subroutine that, given vectors $\veca_1,\ldots,\veca_\ell\in\FF_p^n$ and a degree $1\leq d\leq p-1$, finds a nonzero $\vecu\in\FF_p^\ell$ satisfying $\sum_{i=1}^\ell u_i^d\veca_i=\veczero$. Such a relation lets us combine Fourier samples so that their degree-$d$ coefficient vanishes, and we do this for degrees $d = p-1$ through $2$. Ivanyos and Santha~\cite[Section~3]{IvanyosSantha16} give a deterministic algorithm for this task. Writing $g=\gcd(d,p-1)$, a sufficient number of arbitrary input vectors is $\ell=n^{O(g\log(g+1))} \cdot g^{O(g^2\log(g+1))}$, and the running time is $\poly(\ell,p)$; see~\Cref{thm:ivanyos-santha-diagonal}. This is the subroutine we use.

We observe that a subset sum algorithm also suffices: it finds a nonzero $\vecu\in\{0,1\}^\ell$ with $\sum_i u_i\veca_i=\veczero$. Since $u_i^d=u_i$ for these coefficients, the same relation solves the degree-$d$ problem for every $d$.

Imran and Ivanyos~\cite[Theorem~2]{ImranIvanyos23} give a deterministic subset sum algorithm using $\ell=n^{O(p\log p)} \cdot p^{O(p\log^2p)}$ arbitrary input vectors and $\poly(\ell)$ time. They use it as a subroutine to get a polynomial-time HSP algorithm for nilpotent groups of constant class whose orders have bounded prime factors~\cite[Theorem~1]{ImranIvanyos23}.
The subset sum routine is also used to reduce the problem to a quotient group, however their work does not cover our dihedral groups: $\dih(\ZZ_q^n)$ is not nilpotent when $q$ has an odd prime factor, and for $q=2^k$ its nilpotency class is $k$, which can grow. Their exact theorem also assumes access to a unitary preparing a purification of the subgroup state and its inverse; independent DCP samples do not supply this access, although their central-subgroup conversion itself can use coset-state copies.

Kothari, O'Donnell, and Wu~\cite[Sections~3.2 and~5.2]{KothariODonnellWu26} improve two steps of Imran--Ivanyos's method. For $p\geq5$, their subset sum algorithm for independent and uniformly random inputs requires fewer inputs $\ell=n^{2\lfloor(p+3)/4\rfloor}p^{O(p^2)}$. For $p=3$, Kothari--O'Donnell--Wu gives a deterministic algorithm using $O(n^2)$ arbitrary inputs; see~\Cref{lem:p-ary-subset-sum}.

\subsection{Cryptographic non-implications}
\label{sec:crypto-context}

The equivalence of LWE and suitable forms of EDCP does not make the present algorithm an attack on standard LWE.  The reductions of Brakerski et al.~\cite{brakerski2018learning} produce only a polynomial number of clean EDCP states, while our algorithm uses a quasipolynomial number. An LWE consequence would therefore require either a polynomial-sample version of our algorithm or a reduction that supplies the larger collection of suitably clean states it needs. The same sample barrier was noted for the power-of-two algorithm of Bai et al.~\cite{BaiEtAl25}.

\paragraph{AI disclosure.}
The main result was found by ChatGPT-5.6 Sol in a conversation with the author in July 2026, except for the optimization to use the \cite{IvanyosSantha16} algorithm, which was found in conversation with ChatGPT-6 Astra on 26 September 2026. The author verified and simplified the algorithm, over several conversations with ChatGPT-5.6 Sol, Claude Fable 5.1, ChatGPT-6 Sol and Astra; the author takes full responsibility for correctness, exposition, and attribution.

The above models were also used to help write and edit this manuscript. The introduction and technical overview were primarily written by the author, with AI assistance mainly for copy-editing. In the remaining sections, the author used AI tools over several rounds to expand initial drafts, obtain feedback, and revise the exposition until the author was satisfied with its presentation.

\paragraph{Acknowledgements.} The author thanks Vinod Vaikuntanathan, Noah Shutty, and Seyoon Ragavan for useful discussions. We also thank an anonymous reviewer for pointing out that our quasipolynomial sample and runtime guarantees hold even for smoothness bound $B = \polylog(n)$, not just when $B= O(1)$. The author was supported in part by a Simons investigator award and
NSF grant CNS-2534400, and thanks the UK AISI alignment
project for supporting access to ChatGPT Pro.

\section{Technical Overview}
\label{sec:tech-overview}

We recall that our goal is to solve two-point vector EDCP: we are given samples
\begin{align*}
    \frac{1}{\sqrt{2}}\left(\ket{0}\ket{\vecx_i} + \ket{1}\ket{\vecx_i + \vecs}\right),
\end{align*}
where $\vecs\in\ZZ_q^n$ is a fixed unknown secret and each $\vecx_i\sim \ZZ_q^n$, $i \in [m]$ is uniform and unknown, and our goal is to find $\vecs$. A standard first step is Fourier sampling: measuring the second register of a sample in the $q$-ary Fourier basis returns a uniformly random label $\veca\in\ZZ_q^n$ and leaves, up to global phase, the single-qubit phase state
\begin{align}\label{eqn:phase-qubit}
    \frac{1}{\sqrt{2}}\left(\ket{0} + \omega_q^{\inner{\veca, \vecs}}\ket{1}\right),
\end{align}
where $\omega_q := e^{2\pi i/q}$. We call this a \emph{binary Fourier sample} with label $\veca$; the label $\veca$ is known to the algorithm but the secret $\vecs$ is not. A single phase qubit is of little use on its own, because the phase $\omega_q^{\inner{\veca,\vecs}}$ cannot be read off by a measurement in general. The $q=2$ case is exactly Simon's algorithm~\cite{simon1997power}: performing a Hadamard measurement returns $\inner{\veca,\vecs}\bmod 2$ exactly, and $O(n)$ such equations with independent uniform labels determine $\vecs$.

Now let us think about the case of larger $q$; assume for now that it is even.
If we had gotten lucky, and it so happened that the measurement resulted in $\veca \in (q/2) \cdot \ZZ_q^n \cong \ZZ_2^n$, we would know that $\inner{\veca,\vecs}$ lies in $\{0,q/2\}$, and the same Hadamard-basis measurement would allow us to learn $\left(\frac{2}{q}\inner{\veca,\vecs}\right)\bmod 2$.
The high-level idea of Kuperberg's sieve~\cite{Kuperberg05} is to combine binary Fourier samples so that their labels become progressively ``more divisible'' by $q/2$, until the phases can be read off by a Hadamard measurement. 

The Simon-meets-Kuperberg algorithms~\cite{BonnetainNayaPlasencia18,Csaji21,BaiEtAl25} implement this strategy particularly efficiently for power-of-two moduli, using linear algebra over $\FF_2$ to reduce the modulus by a factor of $2$ at each step. As we will see, this more efficient state-combination procedure does not directly extend to odd primes; our contribution is to overcome this obstacle for smooth moduli.

\subsection{Recalling the $q = 2^k$ algorithm}
\label{sec:power-of-two}
It will be instructive for us to first recall the Simon-meets-Kuperberg algorithms of \cite{BonnetainNayaPlasencia18,BaiEtAl25} for the special case when $q = 2^k$ for some $k \in \NN$. Take $n+1$ phase qubits with labels $\veca_1,\ldots,\veca_{n+1}$ and collect the labels as the rows of a known matrix $\matA\in\ZZ_q^{(n+1)\times n}$. The tensor product of these $n+1$ states is
\begin{align}\label{eqn:polished-power-of-two-combine}
    \frac{1}{\sqrt{2^{n+1}}}
    \sum_{\vecx\in\{0,1\}^{n+1}}
    \omega_q^{\inner{\matA^\top \vecx, \vecs}}\ket{\vecx},
\end{align}
where we interpret $\vecx \in \ZZ_q^{n+1}$ in the phase term. We would like to retain only those $\vecx$ which are in the mod $2$ kernel of $\matA^\top$, i.e., with $\matA^\top\vecx\equiv 0\bmod 2$, since those are exactly the branches whose phase labels lie in $2\ZZ_q^n\cong\ZZ_{q/2}^n$.
A naive hope might be to coherently measure a bit indicating whether $\vecx \in \ker(\matA^\top\bmod2)$ and postselect on success, but the set of such $\vecx$ is far too sparse to do this---there are exactly two such vectors with good probability, namely $\mathbf{0}$ and the nonzero $\vecu \in \FF_2^{n+1}$ kernel vector of $\matA^\top \bmod 2$.

The key observation is that any single \emph{coset} of the kernel is just as good. Unlike the kernel itself, its cosets can be filtered for as follows: we first parameterize $\vecx = (\vecz + t\vecu)$ with $\vecz$ ranging over fixed representatives of the cosets in $\FF_2^{n+1}/\langle\vecu\rangle$ and $t\in\FF_2$ so that \Cref{eqn:polished-power-of-two-combine} reads
\begin{align*}
    \frac{1}{\sqrt{2^{n+1}}}
    \sum_{\substack{\vecz \in \FF_2^{n+1}/\langle\vecu\rangle,\\ t\in\FF_2}}
    \omega_q^{\inner{\matA^\top((\vecz + t\vecu)\bmod2), \vecs}}
    \ket{t}\ket{\vecz}.
\end{align*}
Measuring the second register---that is, measuring which coset of $\langle\vecu\rangle$ the branch lies in---leaves, up to global phase, a binary Fourier sample $\frac{1}{\sqrt2} (\ket0 + \omega_{q/2}^{\inner{\veca',\vecs}}\ket1)$, with uniformly random and known label
\begin{align*}
\veca' := \frac{\matA^\top\bigl(((\vecz+\vecu)\bmod2)-\vecz\bigr)}{2}\in \ZZ_{q/2}^n.
\end{align*}

Iterating this procedure brings the modulus down from $q = 2^k$ to $2$, at a cost of $n+1$ phase qubits per level and hence $n^{O(k)}$ per modulus $2$ phase qubit. A Hadamard measurement at the very end then gives one equation $\inner{\tilde{\veca},\vecs}\equiv c \bmod 2$ with a known uniform $\tilde{\veca}$, and $O(n)$ of them determine $\vecs\bmod 2$ by Gaussian elimination. Once this last bit of the secret is known it can be stripped off and the whole procedure can be repeated on the next bit. Therefore this algorithm has sample and time complexity $n^{O(\log q)}$, which is quasipolynomial when $q=\poly(n)$.

\subsection{The $q = 3^k$ case}
We could try to peel off one factor of $3$ at each level and use the Fourier transform over $\FF_3$ at the end to learn an equation in $\vecs$. But a binary Fourier sample still has only two branches, whereas an exact Fourier measurement modulo $3$ requires a three-branch Fourier state. Moreover, if we measure $\matA^\top\vecx\bmod3$ on $m$ binary samples, the surviving set need not be a coset of the kernel, and need not have three elements.

\paragraph{From qubits to qutrits.}
The mismatch between the support size of the states, which is $2$, and the prime we wish to divide out, which is $3$, motivates converting these binary qubits into qutrits supported on $3$ basis states. Given two Fourier samples with labels $\veca,\veca'$,
\begin{align*}
    \frac{1}{2}
    \left(
        \ket{00}
        + \omega_q^{\inner{\veca', \vecs}}\ket{01}
        + \omega_q^{\inner{\veca, \vecs}}\ket{10}
        + \omega_q^{\inner{\veca + \veca', \vecs}}\ket{11}
    \right),
\end{align*}
we can project the joint state onto $\mathrm{span}(\ket{00},\ket{10},\ket{11})$, which succeeds with probability $3/4$. Relabeling $\ket{00}\mapsto\ket{0}$, $\ket{10}\mapsto\ket{1}$, $\ket{11}\mapsto\ket{2}$, and writing $\vecb := \veca'-\veca$, gives the resulting qutrit
\begin{align*}
    \frac{1}{\sqrt{3}}
    \left(
        \ket{0}
        + \omega_q^{\inner{\veca, \vecs}}\ket{1}
        + \omega_q^{\inner{2\veca + \vecb, \vecs}}\ket{2}
    \right).
\end{align*}
Since $(\veca,\veca')\mapsto(\veca,\vecb)$ is invertible, the labels $\veca, \vecb$ are again independent and uniform. It is convenient to record such a qutrit by a function corresponding to its phase terms $f:\ZZ_3\to\ZZ_q^n$, defined by $f(0) = 0$, $f(1) = \veca$, and $f(2) = 2 \veca + \vecb$. That is,
\begin{align*}
    f(x) \equiv \veca x + \vecb \binom{x}{2} \pmod 3 .
\end{align*}
The above form is the expansion of $f$ in the binomial basis $\binom{x}{1},\binom{x}{2}$, which will be a convenient coordinate system for maps out of $\ZZ_p$ (see \Cref{lem:binomial-coordinates}). Call a qutrit \emph{semi-clean} if its quadratic coefficient vanishes, that is, $\vecb\equiv0\bmod3$. For a semi-clean qutrit the label map is linear modulo $3$, and the binary argument goes through verbatim: combining $n+1$ semi-clean qutrits and measuring the coset of a kernel direction produces a qutrit whose labels are all divisible by $3$, so the modulus drops from $q$ to $q/3$. Therefore the problem has been reduced to manufacturing semi-clean qutrits, or equivalently, clearing the quadratic coefficient $\vecb$.

\paragraph{Clearing the quadratic coefficient $\vecb$ via $\FF_3^n$-subset sum~\cite{ImranIvanyos23}.}
Take $m$ qutrits with phase term functions $f_i(x) = \veca_i x + \vecb_i\binom{x}{2}$.  Their
tensor product is
\begin{align*}
    \frac{1}{\sqrt{3^m}}
    \sum_{\vecx\in\FF_3^m}
    \omega_q^{\inner{g(\vecx), \vecs}}\ket{\vecx},
    \qquad
    g(\vecx) := \sum_{i=1}^m f_i(x_i)
    \equiv
    \sum_{i=1}^m \left(\veca_i x_i + \vecb_i\binom{x_i}{2}\right) \pmod 3 .
\end{align*}
We could try to apply a similar state-combination procedure as the power-of-two algorithm, and for some fixed nonzero vector $\vecu\in\FF_3^m$, we measure the coset of $\vecx$ modulo $\langle\vecu\rangle$. It is not clear yet what $\vecu$ should be; we will see a natural choice shortly.
This measurement leaves a uniform superposition over an affine line $\vecx = \vecz + t\vecu$, $t\in\FF_3$ for some measurement outcome $\vecz$ known to the algorithm. Restricted to that line, $g \bmod 3$ becomes a univariate map of degree at most $2$,
\begin{align*}
    g(\vecz + t\vecu) \equiv \vecc_0 + \vecc_1 t + \vecc_2\binom{t}{2} \pmod 3.
\end{align*}
A simple calculation (see \Cref{lem:affine-binomial-triangularity}) shows that its quadratic coefficient is $\vecc_2 = \sum_{i} u_i^2\vecb_i \bmod 3$.  Because $u_i^2 =
\mathbbm{1}[u_i\neq0]$ in $\FF_3$, this coefficient depends only on the
support of $\vecu$, and it vanishes exactly when
\begin{align*}
    \sum_{i \,:\, u_i \neq 0} \vecb_i \equiv \veczero \pmod 3 .
\end{align*}
Finding such a vector $\vecu$ is an instance of subset-sum over $\FF_3^n$, given input $\vecb_1,\ldots,\vecb_m$. If $m=\Theta(n^2)$, there is a polynomial-time algorithm that finds such a solution, due to \cite{ImranIvanyos23,KothariODonnellWu26}.

One semi-clean qutrit costs $O(n^2)$ qutrits, and one modulus reduction costs $n+1$ semi-clean qutrits, so a single level costs $O(n^3)$ input states. Running $\log_3 q$ levels and collecting $O(n)$ equations per digit gives sample and time complexity $n^{O(\log q)}$, which is quasipolynomial for $q=\poly(n)$.

\subsection{Generalizing to smooth $q$}
The techniques for the $q=3^k$ case can be readily generalized to smooth moduli.

First, we can use a method similar to how we prepared qutrits (or $3$-ary Fourier samples) from binary Fourier samples, to also prepare $p$-ary Fourier samples (and vice versa).

As in the power-of-three algorithm, we reduce these labels modulo $p$ and expand in the binomial basis, giving
\begin{align*}
    f(t) \equiv \sum_{d=1}^{p-1}\binom{t}{d}\veca^{(d)} \pmod p.
\end{align*}
We clear the coefficients from degree $p-1$ down to degree $2$. At degree $d$, collapsing a collection of states to a coset of $\vecu$ changes the degree-$d$ coefficient to
\begin{align}\label{eqn:IS-relation}
    \sum_i u_i^d\veca_i^{(d)}.
\end{align}
Since we want to reduce the degree modulo $p$, we choose $\vecu$ to make this sum zero. Higher degrees do not reappear, so this results in a $p$-ary Fourier sample where the phase terms are described by a degree-$(d-1)$ function modulo $p$.

When $p=3$ and $d=2$, this corresponds exactly to subset sum over $\FF_3^n$. For every $p$ and $d$, a subset-sum solution $\vecu\in\{0,1\}^m$ also suffices, since $u_i^d=u_i$. Such a solution can be supplied by Imran--Ivanyos~\cite{ImranIvanyos23} (or Kothari--O'Donnell--Wu~\cite{KothariODonnellWu26}, but we additionally must prove that the input vectors remain uniformly random). Using the Imran--Ivanyos bound requires $m=n^{O(p\log p)}p^{O(p\log^2p)}$ inputs per degree. The costs multiply over the degrees $p-1,\ldots,2$, so using only this subset sum routine would cost $n^{O(p^2\log p)}p^{O(p^2\log^2p)}$ samples per modulus reduction.

However, we do not need a subset-sum solution $\vecu\in\{0,1\}^m$: any nonzero $\vecu\in\FF_p^m$ making~\Cref{eqn:IS-relation} zero will do. Writing $g=\gcd(d,p-1)$, the Ivanyos--Santha algorithm finds such a relation given
\[
    \ell(p,n,d)
    \ge n^{O(g\log p)}p^{O(g^2\log p)}
\]
inputs. Since $g \ll p$ often, this gives us a better dependence on $B$ in the exponent of $n$:
for each divisor $g\mid(p-1)$, at most $(p-1)/g$ degrees have $\gcd(d,p-1)=g$, since each is a multiple of $g$. Thus the contribution to the exponent of $n$ from each divisor is at most $\frac{p}{g}\,O(g\log p)=O(p\log p)$.
Since $p-1$ has $p^{o(1)}$ divisors by~\Cref{lem:divisor-bound}, the product over all degrees is at most
\[
    n^{O(p^{1+o(1)})}p^{O(p^{2+o(1)})}.
\]

\section{Preliminaries}
\paragraph{Notation and terminology.} We write $\ZZ_q$ to mean $\ZZ/q\ZZ$.

For an abelian group $G$, a subgroup $H\le G$, and $g\in G$, we write $g+H\in G/H$ for the quotient class and $[g]_H\in G$ for a fixed canonical representative of that class. For a function $f$ valued in $G$, we use these notations pointwise: $(f+H)(x)=f(x)+H$ and $[f]_H(x)=[f(x)]_H$.

Throughout this paper, we identify quantum states up to global phase.

\begin{definition}
\label{def:generalized-dihedral}
Let $G$ be a finite abelian group, written additively.
The \textbf{generalized dihedral group} of $G$ is $\dih(G) := G \rtimes_{-1} \ZZ_2$, where the nontrivial element of $\ZZ_2$ acts on $G$ by inversion. That is, writing elements in $\dih(G)$ as pairs $(b,x)\in\ZZ_2\times G$, the group operation is
\begin{align*}
    (b_1,x_1)\cdot(b_2, x_2) = (b_1+b_2,\;x_1+(-1)^{b_1} x_2).
\end{align*}
The dihedral group of $2q$ elements is obtained by taking $G = \ZZ_q$.
\end{definition}
\begin{definition}
    Let $G$ be a finite abelian group, and let $s \in G$ be a fixed secret. The generalized dihedral coset problem asks to recover $s$ from samples of the form
    \begin{align*}
        \frac{1}{\sqrt{2}} \left(\ket{0} \ket{x_i} + \ket{1}\ket{x_i+s}\right),
    \end{align*}
    where $x_i \sim G$ are uniformly random and unknown.
\end{definition}
In this work, we are especially interested in the vector dihedral coset problem, which specializes $G = \ZZ_q^n$.

\subsection{Subset sum over small field size}

\begin{lemma}[Imran--Ivanyos~\cite{ImranIvanyos23} and Kothari--O'Donnell--Wu~\cite{KothariODonnellWu26}]
\label{lem:p-ary-subset-sum}
Let $p$ be prime and $n\geq1$. The following algorithms find a nonempty zero-sum subset of vectors in $\FF_p^n$.
\begin{enumerate}
    \item Theorem~2 of Imran and Ivanyos gives a deterministic algorithm for any $\ell\ge p^{O(p\log^2p)}n^{O(p\log p)}$ input vectors, running in time $\poly(\ell)$.
    \item For $p=3$ and $n\geq2$, Theorem~1.2 of Kothari, O'Donnell, and Wu gives a deterministic algorithm, for any $\ell\geq n^2/3+C n\log n$ input vectors, running in time $\poly(\ell)$, for an absolute constant $C$.
    \item For $p\geq5$, Theorem~4.12 of Kothari, O'Donnell, and Wu gives a deterministic algorithm that succeeds with probability at least $2/3$ on $\ell\ge p^{O(p^2)}n^{2\lfloor(p+3)/4\rfloor}$ independent uniform input vectors and runs in time $\poly(\ell)$.
\end{enumerate}
\end{lemma}
These subset sum routines give alternative ways to find the relations used below. Our algorithm instead uses the following Ivanyos--Santha theorem at every degree.

\begin{theorem}[{Ivanyos--Santha~\cite[Section~3]{IvanyosSantha16}}]
\label{thm:ivanyos-santha-diagonal}
Let $p$ be prime, $n\geq1$, and $1\leq d<p$, and put $g=\gcd(d,p-1)$. Given any $\ell$ vectors $\veca_1,\ldots,\veca_\ell\in\FF_p^n$ with
\begin{align*}
    \ell\geq(n+1)^{g\lceil\log_2(g+1)\rceil}
    g^{\frac{g(g-1)}2\lceil\log_2(g+1)\rceil},
\end{align*}
the deterministic algorithm of Ivanyos and Santha runs in time $\poly(\ell,p)$ and finds a nonzero $\vecu\in\FF_p^\ell$ such that $\sum_{i=1}^\ell u_i^d\veca_i=\veczero$.
\end{theorem}

\subsection{Binomial basis for polynomials}
\begin{lemma}
\label{lem:binomial-coordinates}
Let $p$ be prime.  Every map $g:\ZZ_p\to\ZZ_p^n$ satisfying $g(0)=0$ has a
unique expression
\begin{align*}
    g(x)
    =
    \sum_{d=1}^{p-1}\binom{x}{d}\veca^{(d)},
    \qquad
    \veca^{(d)}
    =
    \sum_{x=0}^{d}(-1)^{d-x}\binom{d}{x}g(x).
\end{align*}
In particular, if $g(1),\ldots,g(p-1)$ are independent and uniform in
$\ZZ_p^n$, then
$\veca^{(1)},\ldots,\veca^{(p-1)}$ are also independent and uniform in
$\ZZ_p^n$.
\end{lemma}

\begin{proof}
The matrix with entries $\binom{x}{d}$ for $x,d=1$ to $p-1$ is lower triangular with every diagonal entry equal to one, and is therefore invertible over $\ZZ_p$. The displayed inverse is the usual finite difference formula. The final assertion follows because an invertible linear map preserves the uniform distribution.
\end{proof}

\begin{lemma}
\label{lem:affine-binomial-triangularity}
Suppose
\begin{align*}
    g(x)
    =
    \sum_{j=1}^{d}\binom{x}{j}\veca^{(j)}
    \pmod p
\end{align*}
for some $d<p$.  Then $g(z+tu)-g(z)$ has degree at most $d$ in $t$, and
its degree-$d$ coefficient in the binomial basis is
$u^d\veca^{(d)}$.  Its linear coefficient is
$u\veca^{(1)}$ plus a value determined by
$\veca^{(2)},\ldots,\veca^{(d)},z$, and $u$.
\end{lemma}

\begin{proof}
The polynomial $\binom{x}{d}$ has degree $d$ and leading coefficient
$1/d!$.  Substituting $x=z+tu$ multiplies its leading coefficient by $u^d$.
Since the degree-$d$ binomial polynomial in $t$ also has leading coefficient
$1/d!$, its coefficient after substitution is $u^d$.  Terms of smaller
degree cannot contribute to it.  Finally, the linear input term contributes
\begin{align*}
    (z+tu)\veca^{(1)}-z\veca^{(1)}
    =tu\veca^{(1)},
\end{align*}
and every other contribution to the linear coefficient comes from the
coefficients of degrees $2$ to $d$.
\end{proof}

\subsection{Group theory and characters}\label{sec:group-theory}

\begin{lemma}\label{thm:fundamental-theorem-finite-abelian}
    Every finite abelian group $G$ is isomorphic to a direct sum of cyclic groups,
    \begin{align*}
        G \cong (\ZZ/ p_1^{\alpha_1} \ZZ)^{n_1} \times \cdots \times (\ZZ/ p_k^{\alpha_k} \ZZ)^{n_k}
    \end{align*}
\end{lemma}

\begin{definition}\label{def:dual-group}
    A character of a finite abelian group $G$ is a homomorphism $\chi:G\to\CC$. The dual group $\hat G$ is the set of these characters, where the group law is pointwise multiplication, i.e.,
    \begin{align*}
        (\chi_1 \chi_2)(g) := \chi_1(g) \cdot \chi_2(g).
    \end{align*}
\end{definition}

\begin{lemma}\label{thm:pontryagin}
    Let $G$ be a finite abelian group. Then, $\hat G \cong G$.
    Further, $\hat{\hat G}$ is canonically isomorphic to $G$, that is, there is an isomorphism $G \to \hat {\hat G}$ that can be defined without fixing a generator.
\end{lemma}
We write this canonical isomorphism as $g \to \chi_g$, for $g \in G$.

\begin{definition}\label{def:annihilator}
For $H\leq G$, its annihilator in $\hat G$ is
\begin{align*}
    H^\perp=\{ \chi \in \hat G:\chi (h)=1\text{ for every }h\in H\}.
\end{align*}
\end{definition}

\begin{lemma}\label{lem:divisor-bound}
Let $N\geq1$ be an integer. Then the number of positive divisors of $N$ satisfies
\[
\#\{g\geq1:g\mid N\}=N^{o(1)}.
\]
\end{lemma}

\begin{proof}
We will show that for every fixed constant $c > 0$, we have that
\begin{align*}
    \#\{g\geq1:g\mid N\} = O(N^c)\qquad(N\to\infty).
\end{align*}
Let $c > 0$ be an arbitrary constant.
Write $N=\prod_{i=1}^k p_i^{\alpha_i}$, where the $p_i$ are distinct primes and $\alpha_i\geq1$.
Then, we have that
\[
\#\{g\geq1:g\mid N\}=\prod_{i=1}^k(\alpha_i+1).
\]
If $p_i^c\geq2$, then $\alpha_i+1\leq2^{\alpha_i}\leq p_i^{c\alpha_i}$. Only finitely many primes satisfy $p_i^c<2$, and for each of them
\[
\sup_{\alpha_i\geq1}(\alpha_i+1)p_i^{-c\alpha_i}<\infty
\]
because $p_i^c>1$. Multiplying these bounds gives $\prod_{i=1}^k(\alpha_i+1)=O(N^c)$.
\end{proof}

\section{Our algorithm}\label{sec:main-algo}

In this section, we present our algorithm and its proof of correctness for the dihedral coset problem over $\operatorname{Dih}(G)$, for a general abelian group $G$ equipped with the necessary computational subroutines. We later specialize to the case $G = (\ZZ/ q\ZZ)^n$. 

\paragraph{Black-box group model.} We assume that we have black-box access to the group, i.e., every group element is identified by a unique string (of length $O(\log |G|)$, say), and we are given access to an oracle that performs the group operation. Further, we are given a set of generators for $G$. Cheung and Mosca~\cite{CheungMosca01} give an algorithm that, given black-box access to $G$ as described above, computes a cyclic decomposition of $G$ as in~\Cref{thm:fundamental-theorem-finite-abelian}. Therefore, we assume without loss of generality that we have efficient algorithms that compute the forward and inverse maps for the cyclic decomposition isomorphism
\begin{align*}
    G \cong (\ZZ/ p_1^{\alpha_1} \ZZ)^{n_1} \times \cdots \times (\ZZ/ p_k^{\alpha_k} \ZZ)^{n_k}.
\end{align*}
This allows us to work with quotient groups and characters of the group easily, so for our purposes, we are effectively working with the explicit cyclic decomposition group on the right hand side of the above equation. Nevertheless, we will write the rest of this section in terms of general groups to avoid keeping track of multiple indices, and keep the notation as simple as possible. For the relevant group theory background, please see \Cref{sec:group-theory}, and for a more elementary presentation, please refer to \Cref{sec:tech-overview}.

\begin{notation}
    For a prime $p$, a finite abelian group $G$, $s \in G$, and function $f: \FF_p \to \hat G$, we define
    \begin{align*}
        \ket{\phi^{f, s}_G} := \frac{1}{\sqrt{p}} \sum_{x \in \FF_p} \chi_s(f(x)) \ket{x}.
    \end{align*}
    Here $\chi_s$ is a character of $\hat{G}$. We call $\left(f, \ket{\phi^{f, s}_G}\right)$ a $p$-ary Fourier sample. If $p=2$, we will ignore global phase, and also call $\left(f(1) - f(0), \frac{\ket{0} + \chi_s(f(1) - f(0)) \ket{1}}{\sqrt{2}}\right)$ a binary Fourier sample.
\end{notation}
When $G = \ZZ_q^n$, the character $\chi_s$ can be chosen to be $\chi_\vecs(\veca) = \omega_q^{\inner{\vecs, \veca}}$ for every $\veca \in \ZZ_q^n$.

\subsection{Preparing Fourier samples}

\begin{lemma}\label{lem:edcp-fourier-sampling}
    Let $G$ be a finite abelian group and let $s, x \in G$ be arbitrary. Applying a Fourier transform over $G$ to the second register of the quantum state $\frac{1}{\sqrt{2}} (\ket{0, x} + \ket{1, x+s})$ and measuring it gives a uniformly random Fourier label $a \in \hat{G}$ as the measurement outcome and the corresponding binary Fourier sample
    \begin{align}\label{eqn:fourier-sample}
        \frac{1}{\sqrt{2}} (\ket{0} + \chi_s(a) \ket{1}),
    \end{align}
    as the residual quantum state.
\end{lemma}

\begin{proof}
Use the Fourier transform convention $\ket{y}\mapsto |G|^{-1/2}\sum_{a\in\hat G}\chi_y(a)\ket{a}$. Since $\chi_{x+s}(a)=\chi_x(a)\chi_s(a)$, applying this transform to the second register gives
\begin{align*}
    \frac{1}{\sqrt{2|G|}}\sum_{a\in\hat G}\chi_x(a)\bigl(\ket{0}+\chi_s(a)\ket{1}\bigr)\ket{a}.
\end{align*}
Thus every label is measured with probability $1/|G|$. After measurement, the factor $\chi_x(a)$ is a global phase of the qubit and can be ignored, leaving the state in~\eqref{eqn:fourier-sample}.
\end{proof}

\begin{lemma}\label{lem:binary-to-p-ary-fourier-sample}
    Let $p$ be a prime, $H$ be a finite abelian group, and fix $s \in H$.
    There is a quantum procedure that takes $O(p)$ binary Fourier samples of the form $\left(a, \frac{\ket{0} + \chi_s(a) \ket{1}}{\sqrt{2}}\right)$ and outputs a $p$-ary Fourier sample $\left(f, \ket{\phi_H^{f, s}} \right)$, for some $f: \FF_p \to \hat H$, $f(0) = 0$.
    If the binary Fourier labels are i.i.d.~uniformly random, $f$ is a uniformly random function. The procedure runs in time polynomial in $p$ and the group-operation cost, and succeeds with probability $\Omega(1)$.
\end{lemma}

\begin{proof}
Collect $\ell$ binary Fourier samples, with labels $a_1,\ldots,a_\ell$. Their joint state is
\[
    2^{-\ell/2}\sum_{\vecb \in \{0,1\}^\ell} \chi_s \left(\sum_{i=1}^\ell b_i a_i \right) \ket{\vecb}.
\]
We will partition $\{0,1\}^\ell$ into components of size $\ell+1$, measure the component label, and retain $p$ strings from that component.

It is tempting to take $\ell=\lceil\log_2 p\rceil$ and take a single component corresponding to the binary encodings of $0,\ldots,p-1$. This succeeds with probability $p/2^\ell\geq 1/2$, but the resulting values of $f$ are sums of only $\ell$ random input labels and therefore cannot be independently uniformly random. Instead, we will take $\ell = 2^{\lceil\log_2 p\rceil}-1$, and our partition will be such that each component consists of a ``central'' bit string $\vecc$ and all the bit strings of Hamming distance $1$ from it, i.e., $\vecc+\mathbbm{1}_i$, for $i\in[\ell]$. Relative to the center's phase, the labels on these $\ell$ neighbors are independently uniform.

Our centers will be the vectors in the right kernel of the matrix $\matM \in \FF_2^{\lceil \log_2 p \rceil \times \ell}$ whose columns are all the nonzero vectors of length $\lceil \log_2 p \rceil$. The image of every other string is nonzero and therefore matches exactly one column; flipping the bit corresponding to that column changes the result to zero. Thus every string is either a center or has Hamming distance $1$ from a unique center. Therefore, we can collapse the state to a single component by measuring the nearest center $\vecc$, and then we can relabel $\vecc \mapsto 0$ and $\vecc + \mathbbm{1}_i \mapsto i$. The resulting state is a uniform superposition over $\ell+1$ branches, which we then project to labels in $\FF_p$.
This succeeds with probability $p/(\ell+1) \ge 1/2$, resulting in the state, up to global phase, $\frac{1}{\sqrt{p}}\sum_{x \in \FF_p} \chi_s(f(x)) \ket{x}$, where $f(x)=(-1)^{c_x}a_x$ for $x\ne0$, and $f(0)=0$. Since the $a_i$'s are uniformly random, we have that $f$ is uniformly random among functions with $f(0) = 0$. The success probability does not depend on the input labels or $s$: every branch has the same magnitude. Thus failed attempts can be discarded and replaced with fresh inputs, and the expected number of binary samples per output is $\ell(\ell+1)/p<4p$.
\end{proof}

\begin{lemma}\label{lem:p-ary-to-binary-fourier-sample}
   Let $p$ be a prime, $H$ be a finite abelian group, and fix $s \in H$.
   There is a quantum procedure that takes a $p$-ary Fourier sample $\left(f, \ket{\phi_H^{f, s}} \right)$ and outputs a binary Fourier sample $\left(a, \frac{\ket{0} + \chi_s(a) \ket{1}}{\sqrt{2}}\right)$.
   If the function $f$ is uniformly random conditioned on $f(0) = 0$, then $a$ is uniformly random. The procedure runs in time polynomial in $p$ and the group-operation cost, and succeeds with probability $\ge 2/3$.
\end{lemma}

\begin{proof}
Partition $\FF_p$ into fixed pairs, for instance $\{(0,1), (2, 3), \ldots\}$, leaving one index unpaired if $p$ is odd. We measure the pair index to collapse the state to be supported only on a single pair. (We declare failure if the outcome is the unpaired index.)
This succeeds with probability $1$ for $p=2$ and $(p-1)/p\geq 2/3$ for odd $p$, independently of $f$ and $s$. If the measured pair is $\{x,y\}$, we can relabel $x \mapsto 0, y \mapsto 1$. After ignoring the common phase $\chi_s(f(x))$, the normalized state is $\frac{\ket{0}+\chi_s(f(y)-f(x))\ket{1}}{\sqrt{2}}$. The procedure outputs the classical label $a=f(y)-f(x)$; and when $f$ is uniform (subject to $f(0)=0$), $f(y)-f(x)$ is uniform as well.
\end{proof}

\subsection{A state combination procedure}\label{sec:quotienting}

\begin{lemma}\label{lem:quotienting-correctness}
    Let $H$ be a finite abelian group. Given $\ell$ $p$-ary Fourier samples with fixed secret $s \in H$ and arbitrary known labels $f_1, \ldots, f_\ell : \FF_p \to \hat H$, and a nonzero vector $\vecu \in \FF_p^\ell$, there is a quantum procedure that measures a uniformly random coset of $\langle\vecu\rangle\leq\FF_p^\ell$ and outputs a $p$-ary Fourier sample with label
    \begin{equation}\label{eq:affine-output-table}
        g(t)=\sum_{i=1}^\ell\bigl(f_i(z_i+t u_i)-f_i(z_i)\bigr),
    \end{equation}
    along with a representative $\vecz$ of the measured coset. The representative is chosen by any fixed efficient rule that depends only on $\vecu$ and the coset. The procedure runs in time $\poly(\ell, p, \log |H|)$.
\end{lemma}

\begin{proof}
Given the joint input state $\frac{1}{p^{\ell/2}}\sum_{\vecx\in\FF_p^\ell}\chi_s\!\left(\sum_{i=1}^\ell f_i(x_i)\right)\ket{\vecx}$, we coherently compute and measure which coset of $\langle\vecu\rangle$ contains the index vector $\vecx$, collapsing to a superposition on that coset. Each coset contains $p$ orthogonal basis vectors of magnitude $p^{-\ell/2}$, so each coset occurs with probability $p^{1-\ell}$. Every vector in the measured coset is of the form $\vecz+t\vecu$ for a unique $t$, and the remaining state is
\begin{align*}
    \frac{1}{\sqrt p}\sum_{t\in\FF_p}\chi_s\!\left(\sum_{i=1}^\ell f_i(z_i+t u_i)\right)\ket{\vecz+t\vecu}.
\end{align*}
We then reversibly relabel $\ket{\vecz+t\vecu} \mapsto \ket t$, so that the relative phase of $\ket{t}$ with respect to $\ket{0}$ is $\chi_s(g(t))$ in~\eqref{eq:affine-output-table}, with $g(0)=0$. Computing the coset label, relabeling its $p$ vectors, and computing the truth table of $g$ take time polynomial in $\ell$, $p$, and $\log|H|$.
\end{proof}

\begin{lemma}\label{lem:uniform-affine-table-combination}
    Let $f_1, \ldots, f_\ell, g$ be functions as in \Cref{lem:quotienting-correctness}.
    If $f_1, \ldots, f_\ell$ are i.i.d.~uniformly random functions that vanish on $0$, and $\vecu$ is chosen independently, then $g$ is a uniformly random function that vanishes on $0$, even conditioned on $\vecu$ and $\vecz$.
\end{lemma}

\begin{proof}
It is clear that $g(0) = 0$.
Fix $\vecu$ and $\vecz$. The choice of $\vecu$ is independent of the input functions, and by~\Cref{lem:quotienting-correctness}, $\vecz$ is uniform for every fixed choice of those functions. Thus the posterior on $f_i$ remains independent and uniform conditioned on $\vecu, \vecz$. Choose an index $j$ with $u_j\ne0$, and fix the other input functions. The contribution of $f_j$ to $g$ is the function $t\mapsto f_j(z_j+t u_j)-f_j(z_j)$, which vanishes at $0$.

As $t$ ranges over $\FF_p$, the argument $z_j+t u_j$ ranges over all of $\FF_p$, and therefore this contribution is uniform. The other functions $f_i, i \neq j$ contribute a fixed function that also vanishes at $0$, and this preserves uniformity.
\end{proof}

\begin{lemma}\label{lem:kernel-uniformity}
    Let $f_1, \ldots, f_\ell, g$ be functions as in \Cref{lem:quotienting-correctness}, with $f_i (0) = 0$. Let $A \le \hat H$, and choose canonical representatives in $\hat H$ for each coset of $A$, with $0$ representing $A$. Suppose that the vector $\vecu$ in \Cref{lem:quotienting-correctness} depends only on the $[f_i]_{A}$ functions. Suppose further that $f_i - [f_i]_{A}$ are i.i.d.~uniformly random $A$-valued functions that vanish at $0$, even conditioned on the entire collection $[f_1]_{A}, \ldots, [f_\ell]_{A}$. Then, $g - [g]_{A}$ is a uniformly random $A$-valued function that vanishes at $0$, even conditioned on $[g]_{A}$ and $\vecz$. Here $[f_i]_{A}$ and $[g]_{A}$ apply the chosen representatives pointwise.
\end{lemma}

\begin{proof}
Fix the representative functions $[f_1]_{A},\ldots,[f_\ell]_{A}$. This fixes $\vecu$, while the functions $f_i-[f_i]_{A}$ remain independent and uniform by assumption. The coset measurement of~\Cref{lem:quotienting-correctness} is uniform even when all the input functions are fixed. We may therefore also condition on its outcome $\vecz$ without changing the distribution of these $A$-valued functions. Substituting $f_i=[f_i]_{A}+(f_i-[f_i]_{A})$ into~\eqref{eq:affine-output-table}, we have
\begin{align*}
g(t)-[g(t)]_{A}
&=\sum_{i=1}^\ell\bigl((f_i-[f_i]_{A})(z_i+t u_i)-(f_i-[f_i]_{A})(z_i)\bigr)\\
&\quad+\sum_{i=1}^\ell\bigl([f_i(z_i+t u_i)]_{A}-[f_i(z_i)]_{A}\bigr)-[g(t)]_{A}.
\end{align*}
The terms involving $[f_i]_{A}$ determine $g+A$, and hence determine $[g]_{A}$.
Applying~\Cref{lem:uniform-affine-table-combination} to the $A$-valued functions $f_i - [f_i]_{A}$ shows that the first sum is uniform among $A$-valued functions vanishing at $0$. The remaining terms form a fixed $A$-valued function: the second sum and $[g(t)]_{A}$ represent the same coset, and both vanish at $t=0$. Adding this fixed function to the first sum preserves uniformity.
\end{proof}

\subsection{Modulus reduction}\label{sec:modulus-reduction}

Whenever $\hat H/A\cong\FF_p^r$ below, we fix such an isomorphism and use it to view $a+A$ as a vector in $\FF_p^r$.

\begin{lemma}\label{lem:degree-cancellation}
Let $f_1, \ldots, f_\ell, g : \FF_p \to \hat H$ and $\vecu$ be as in \Cref{lem:quotienting-correctness}. Suppose that $\hat H / A \cong \FF_p^r$ for some $A \le \hat H$, and suppose $f_i(x)+A = \sum_{j=1}^d \binom{x}{j} \veca_i^{(j)}$ for some degree $1 \le d < p$ and coefficients $\veca_i^{(j)} \in \FF_p^r$. If $\vecu$ satisfies
\begin{align}\label{eq:degree-relation}
    \sum_{i=1}^\ell u_i^{d} \veca_i^{(d)} = 0,
\end{align}
then $g+A$ has degree at most $d-1$. That is, there exist $\vecb^{(1)}, \ldots, \vecb^{(d-1)} \in \FF_p^r$ such that $g(x)+A = \sum_{j=1}^{d-1} \binom{x}{j} \vecb^{(j)}$.
\end{lemma}

\begin{proof}
By~\eqref{eq:affine-output-table}, in $\FF_p^r$ we have
\begin{align*}
g(t)+A=\sum_{i=1}^\ell\sum_{j=1}^d\left(\binom{z_i+t u_i}{j}-\binom{z_i}{j}\right)\veca_i^{(j)}.
\end{align*}
By~\Cref{lem:affine-binomial-triangularity}, the coefficient of $\binom{t}{d}$ is $\sum_i u_i^d\veca_i^{(d)}=0$ by~\eqref{eq:degree-relation}. Thus $g(t)+A$ has degree at most $d-1$. Also $g(0)+A=0$, so its constant term vanishes. For $d=1$, this gives $g(t)+A=0$.
\end{proof}

\begin{lemma}\label{lem:coefficient-uniformity}
    Consider the setting of \Cref{lem:degree-cancellation}. Suppose that conditioned on degree-$d$ coefficients $\veca_1^{(d)}, \ldots, \veca_\ell^{(d)}$, the lower degree coefficients $\{\veca_i^{(j)}\}_{i \in [\ell], j \in [d-1]}$ are jointly i.i.d. uniformly randomly distributed in $\FF_p^r$. Suppose further that $\vecu$ is a (possibly randomized) function of the degree-$d$ coefficients $\veca_1^{(d)}, \ldots, \veca_\ell^{(d)}$, and independent of the lower degree coefficients. Then, the output coefficients $\vecb^{(1)}, \ldots, \vecb^{(d-1)}$ are jointly i.i.d.~uniform, even conditioned on $\veca_1^{(d)}, \ldots, \veca_\ell^{(d)}$.
\end{lemma}

\begin{proof}
The statement is vacuous for $d=1$, so suppose $d\ge2$. Fix the degree-$d$ coefficients and the independent coins used to choose $\vecu$. This fixes a nonzero $\vecu$. We may also fix the measured coset representative $\vecz$: by~\Cref{lem:quotienting-correctness}, its distribution is uniform for every fixed collection of input tables, so this conditioning does not change the distribution of their lower coefficients.

Choose $i$ with $u_i\ne0$ and fix the lower coefficients of every other input. For $j<d$, the contribution of $\veca_i^{(j)}$ to $g(t)+A$ is
\begin{align*}
\left(\binom{z_i+t u_i}{j}-\binom{z_i}{j}\right)\veca_i^{(j)}.
\end{align*}
By~\Cref{lem:affine-binomial-triangularity}, its coefficient of $\binom{t}{j}$ is $u_i^j\veca_i^{(j)}$, and it has no terms of higher degree. The fixed degree-$d$ terms cancel at degree $d$ by~\eqref{eq:degree-relation}. Thus $\vecb^{(j)}$ depends on $\veca_i^{(j)},\ldots,\veca_i^{(d-1)}$, but not on any lower coefficient of the $i$th input.

At degree $d-1$, every contribution to $\vecb^{(d-1)}$ except $u_i^{d-1}\veca_i^{(d-1)}$ is fixed. Since $u_i^{d-1}\ne0$, this output coefficient is uniform, and fixing it also fixes $\veca_i^{(d-1)}$. Now suppose the output coefficients of degrees $j+1,\ldots,d-1$ have been fixed. Working downward in the same way, they fix $\veca_i^{(j+1)},\ldots,\veca_i^{(d-1)}$. The only unfixed contribution to $\vecb^{(j)}$ is then $u_i^j\veca_i^{(j)}$. Since $\veca_i^{(j)}$ is still independent and uniform and $u_i^j\ne0$, $\vecb^{(j)}$ is uniform even conditioned on all higher output coefficients. Continuing to $j=1$ proves joint independence and uniformity.
\end{proof}

\begin{lemma}[Degree-relation routine]
\label{lem:degree-relation-routine}
Let $p$ be prime, $r\geq1$, and $1\leq d<p$, and put $g=\gcd(d,p-1)$. Set
\[
    \ell(p,r,d)=(r+1)^{g\lceil\log_2(g+1)\rceil}
    g^{\frac{g(g-1)}2\lceil\log_2(g+1)\rceil}.
\]
The deterministic Ivanyos--Santha algorithm, given any $\ell=\ell(p,r,d)$ vectors $\veca_1,\ldots,\veca_\ell\in\FF_p^r$, runs in time at most $T(p,r,d)=\poly(\ell(p,r,d),p)$ and finds a nonzero $\vecu\in\FF_p^\ell$ satisfying
\begin{align*}
    \sum_{i=1}^\ell u_i^d\veca_i=\veczero.
\end{align*}
In particular, $\ell(p,r,1)=r+1$.
\end{lemma}

\begin{proof}
Apply~\Cref{thm:ivanyos-santha-diagonal} with $n=r$. The chosen number of vectors meets its hypothesis, giving the required relation and running time. When $d=1$, we have $g=1$ and the formula gives $r+1$ vectors. In the degree-reduction procedure, the algorithm inspects only the supplied degree-$d$ coefficient vectors.
\end{proof}

Below, $\ell(p,r,d)$ and $T(p,r,d)$ refer to the routine in~\Cref{lem:degree-relation-routine}.

\begin{lemma}\label{lem:degree-reduction-to-linear}
Let $p$ be prime, let $r\geq1$ be an integer, $H$ be a finite abelian group, and let $A \le \hat H$ with $\hat H/A\cong\FF_p^r$. There is a Las Vegas quantum procedure that takes as input $p$-ary Fourier samples $\left(f_i , \ket{\phi_H^{f_i, s}}\right)$ over $H$ with fixed secret $s$ and i.i.d.~uniformly random functions $f_i: \FF_p\to\hat H$ that vanish at $0$, and outputs a $p$-ary Fourier sample $\left(g,\ket{\phi_H^{g,s}}\right)$ with
\begin{align*}
    g(t)+A=t\veca
\end{align*}
for all $t \in \FF_p$, where $\veca \sim \FF_p^r$ is uniformly random. Further, we have that $g(0) = 0$, and conditioned on $g+A$, the values of $g(t)$ for $t \neq 0$ are independently uniform in the corresponding coset of $A$. The expected number of input samples and time required by the procedure are at most $m$ and $T'$ respectively, where
\begin{align*}
    m = \prod_{d=2}^{p-1}\ell(p,r,d), \qquad \text{ and } \qquad T' = O\!\left(p\prod_{d=2}^{p-1}\left(T(p,r,d)+\poly\!\left(\ell(p,r,d),p,r,\log|H|\right)\right)\right).
\end{align*}
\end{lemma}

\begin{proof}
A degree-$d$ pass takes samples whose quotient coefficients of degrees $d+1,\ldots,p-1$ have already been cleared, and clears their degree-$d$ coefficients. Begin with a supply of fresh original input samples. For $d=p-1,p-2,\ldots,2$, the degree-$d$ pass does the following:
\begin{enumerate}
\item Draw $\ell: = \ell(p,r,d)$ independent samples $(f_i, \ket{\phi^{f_i, s}_H}), i \in [\ell]$ from the previous pass, and compute the degree-$d$ coefficient $\veca_i^{(d)}$ of the functions $f_i$ modulo $A$:
\begin{align*}
    \veca_i^{(d)}=\sum_{x=0}^{d}(-1)^{d-x}\binom{d}{x}\bigl(f_i(x)+A\bigr).
\end{align*}
\item Run the degree-$d$ relation routine of~\Cref{lem:degree-relation-routine} on these degree-$d$ coefficients $\veca_1^{(d)}, \ldots, \veca_\ell^{(d)}$ to get $\vecu$ such that \eqref{eq:degree-relation} holds.
\item Apply~\Cref{lem:quotienting-correctness} to the block using $\vecu$, and provide its output as one sample from the new supply.
\end{enumerate}
Return one sample from the final degree pass.
For each $t\ne0$, the pair $\bigl(f_i(t)+A,\ f_i(t)-[f_i(t)]_{A}\bigr)$ is uniform in $\FF_p^r\times A$. These pairs are independent across $i$ and $t$. Thus the $A$-valued parts are uniform even after the functions $f_i$ are fixed modulo $A$, and~\Cref{lem:binomial-coordinates} gives
\begin{align*}
f_i(t)+A=\sum_{j=1}^{p-1}\binom{t}{j}\veca_i^{(j)},
\end{align*}
where the $\veca_i^{(j)}$ are independent uniform vectors in $\FF_p^r$. Suppose the samples entering a degree-$d$ pass have
\begin{align*}
f_i(t)+A=\sum_{j=1}^d\binom{t}{j}\veca_i^{(j)},
\end{align*}
where the $\veca_i^{(j)}$ are independent uniform. The relation returned by the deterministic routine and \Cref{lem:quotienting-correctness,lem:degree-cancellation} give an output sample with
\begin{align*}
g(t)+A=\sum_{j=1}^{d-1}\binom{t}{j}\vecb^{(j)}.
\end{align*}
The lower output coefficients $\vecb^{(1)}, \ldots, \vecb^{(d-1)}$ remain uniform, by \Cref{lem:coefficient-uniformity}. 
Conditional on $f_i + A$, the functions $f_i-[f_i]_{A}$ remain independent uniform. Since $\vecu$ depends only on the functions $f_i + A$, by \Cref{lem:kernel-uniformity}, we have that $g-[g]_{A}$ is uniform even after $g+A$ is fixed. Thus the next pass receives the same uniformity properties.

After the final pass, $g(t)+A=t\veca$ with uniform $\veca\in\FF_p^r$, and conditional on $g + A$, the values $g(t)-[g(t)]_{A}$ for $t\ne0$ are independent uniform in $A$, as required. For $p=2$, the returned input sample already has these properties.

Each degree-$d$ pass consumes exactly $\ell(p,r,d)$ outputs from the preceding pass, because its relation routine always returns a valid direction. Multiplying these costs gives $m$ original input samples per output. The additional time per pass is at most $T(p,r,d)$ plus a polynomial in $\ell(p,r,d)$, $p$, $r$, and $\log|H|$ for coefficient computation and state combination. Choose the polynomial term to dominate $\ell(p,r,d)$. The time recurrence then has fewer than $p$ degree levels, and summing their costs gives the displayed bound on $T'$. For $p=2$, both products are empty, so the returned input also meets the bounds.
\end{proof}

\begin{lemma}\label{lem:modulus-reduction-correctness}
    Let $p$ be a prime, let $r \ge 1$ be an integer, let $H$ be a finite abelian group with $A \le \hat H$ such that $\hat H / A \cong \FF_p^r$. Let $K := A^\perp \cong \FF_p^r$. Then, there exists a Las Vegas quantum algorithm that takes as input $p$-ary Fourier samples $\left( f_i,  \ket{\phi^{f_i, s}_H} \right)$ over $H$ with fixed secret $s \in H$ and i.i.d.~uniformly random functions $f_i : \FF_p \to \hat H$ that vanish at $0$, and outputs a $p$-ary Fourier sample $\left( g,  \ket{\phi^{g, s + K}_{H/K}} \right)$ over $H/K$. Further, $g: \FF_p \to A$ is a uniformly random function that vanishes at $0$.
    This procedure uses at most $m$ input samples in expectation, and runs in time at most $T'$ in expectation, where
    \begin{align*}
        m = \prod_{d=1}^{p-1}\ell(p, r, d), \qquad \text{and } \qquad T' = O\!\left(
p\prod_{d=1}^{p-1}
\left(T(p,r,d)+\poly\!\left(\ell(p,r,d),p,r,\log |H|\right)\right)
\right).
    \end{align*}
\end{lemma}

\begin{proof}
Use~\Cref{lem:degree-reduction-to-linear} on disjoint fresh inputs to obtain independent samples $\left(f_i, \ket{\phi^{f_i, s}_H}\right)$ for which $f_i(t) + A = t\veca_i$, with the $\veca_i$ independent and uniform in $\FF_p^r$. Collect $\ell(p,r,1)$ such samples and run the degree-$1$ routine from \Cref{lem:degree-relation-routine} on these vectors. Combining the states using~\Cref{lem:quotienting-correctness} gives a $p$-ary Fourier sample $\left(g, \ket{\phi^{g, s}_H}\right)$ such that
\begin{align*}
g(t)+A=t\sum_{i=1}^\ell u_i\veca_i=0.
\end{align*}
Since $g(t)+A=A$ and $0$ represents $A$, we have $[g]_{A}=0$ and hence $g-[g]_{A}=g$.
By~\Cref{lem:kernel-uniformity}, we have that $g$ is uniformly random, conditioned on $g(0) = 0$. For every $a\in A$ and $k\in K=A^\perp$, we have $\chi_{s+k}(a)=\chi_s(a)\chi_k(a)=\chi_s(a)$. Under $A\cong\widehat{H/K}$, this gives $\ket{\phi_H^{g,s}}=\ket{\phi_{H/K}^{g,s+K}}$.

A degree-$1$ pass uses $\ell(p,r,1)$ outputs of~\Cref{lem:degree-reduction-to-linear}, giving the stated sample bound $m$. The relation routine and final state combination add $T(p,r,1)+\poly(\ell(p,r,1),p,r,\log|H|)$ time. Applying the same time recurrence as in~\Cref{lem:degree-reduction-to-linear} and absorbing the fewer than $p$ degree levels gives the stated bound $T'$.
\end{proof}

\subsection{Terminal recovery}

\begin{lemma}\label{lem:terminal-recovery}
Let $p$ be prime, let $r\geq1$ be an integer, and let $H$ be a finite abelian group with fixed coordinates $\hat H\cong\FF_p^r$. For $a\in\hat H$, write $\veca\in\FF_p^r$ for its coordinates; the unknown $s\in H$ is represented by $\vecs\in\FF_p^r$ through $\chi_s(a)=\omega_p^{\inner{\veca,\vecs}}$. There is a Las Vegas quantum procedure that takes independent $p$-ary Fourier samples $\left(f_i,\ket{\phi_H^{f_i,s}}\right)$ over $H$, where the functions $f_i:\FF_p\to\hat H$ are i.i.d.~uniform and vanish at $0$, and returns $\vecs$ with probability $1$. Its expected number of input samples is at most $m$, and its expected running time is at most $T'$, where
\begin{align*}
    m = (r+2)\prod_{d=2}^{p-1}\ell(p,r,d), \qquad \text{ and } \qquad
    T' = O\!\left(m\,\poly\!\left(p,r,\log|H|,\sum_{d=2}^{p-1}\bigl(\ell(p,r,d)+T(p,r,d)\bigr)\right)\right).
\end{align*}
\end{lemma}

\begin{proof}
Apply~\Cref{lem:degree-reduction-to-linear} with $A=\{0\}$ to fresh, disjoint inputs. Each call returns a sample with $g(t)=t\veca$, where $\veca$ is uniform in $\FF_p^r$. Its state is
\begin{align*}
\ket{\phi_H^{g,s}}=\frac{1}{\sqrt p}\sum_{t\in\FF_p}\omega_p^{t\inner{\veca,\vecs}}\ket t.
\end{align*}
An inverse Fourier transform over $\FF_p$ followed by a measurement gives $\inner{\veca,\vecs}$ exactly. Calls on disjoint inputs therefore give independent uniform vectors $\veca$ and their corresponding linear equations in $\vecs$. Continue until the obtained vectors $\veca$ span $\FF_p^r$, then solve the equations by Gaussian elimination.

When their current span has dimension $j$, a fresh uniform vector lies outside it with probability $1-p^{j-r}$. Thus the expected number of calls is
\begin{align*}
\sum_{j=0}^{r-1}\frac{1}{1-p^{j-r}}
=r+\sum_{j=1}^{r}\frac{1}{p^j-1}<r+2,
\end{align*}
since $1/(p^j-1)\leq 2^{1-j}$ and $\sum_{j=1}^r2^{1-j}<2$, giving the desired upper bound on the expected sample complexity. 

Since each original sample is used in at most one attempt per degree and each final equation uses at least one original sample, the total work is at most the number of original samples times a polynomial in $p,r,\log|H|$ and $\sum_{d=2}^{p-1}(\ell(p,r,d)+T(p,r,d))$, giving the stated expected time bound.
\end{proof}

\begin{lemma}\label{lem:character-correction}
    Let $p$ be prime, let $H$ be a finite abelian group with fixed secret $s\in H$, and let $A\leq\hat H$. Let $t\in H$ be such that $\chi_t$ is a known, efficiently evaluatable extension of $\chi_s|_A$. Define $K := A ^\perp \le H$, so that $\hat K \cong \hat H / A$.
    Then a deterministic quantum procedure takes as input a $p$-ary Fourier sample $\left(f,\ket{\phi_H^{f,s}}\right)$ with $f(0)=0$ and $\chi_s |_{A}$, multiplies the phase of $\ket{x}$ by $\chi_t(f(x))^{-1}$, and outputs a $p$-ary Fourier sample $\left(g, \ket{\phi_{K}^{g, r}}\right)$ over $K$, where $g: \FF_p \to \hat H/ A$,
    \begin{align*}
        r = s - t \in K, \qquad \text{ and } \qquad g(x) = f(x) + A.
    \end{align*}
    If $f$ is a uniformly random function such that $f(0) = 0$, so is $g$.
\end{lemma}

\begin{proof}
Since $\chi_t$ extends $\chi_s|_{A}$, for every $h\in A$ we have $\chi_{s-t}(h)=\chi_s(h)\chi_t(h)^{-1}=1$. Hence $r=s-t$ belongs to $K=A^\perp$. In particular, $\chi_r$ is trivial on $A$, so its value at $f(x)$ depends only on the coset $g(x)=f(x)+A$. The prescribed phase correction therefore changes the state to
\begin{align*}
\frac{1}{\sqrt p}\sum_{x\in\FF_p}\chi_s(f(x))\chi_t(f(x))^{-1}\ket{x}
=\frac{1}{\sqrt p}\sum_{x\in\FF_p}\chi_r(g(x))\ket{x}
=\ket{\phi_K^{g,r}},
\end{align*}
using $\hat H/A\cong\hat K$.
Since $f$ is uniform subject to $f(0)=0$, the values of $g(x)=f(x)+A$ for $x\ne0$ are independent and uniform in $\hat H/A$.
\end{proof}

\subsection{Putting everything together}
\label{sec:main-theorem-proof}
We are now ready to describe the main algorithm.

\begin{algorithm}\label{algo:main}
As input, we are given black-box access to a finite abelian group $G$, and independent coset states $\frac{1}{\sqrt{2}}\bigl(\ket{0,x}+\ket{1,x+s}\bigr)$, where each $x$ is independent and uniform in $G$ and $s\in G$ is fixed and unknown. Use the Ivanyos--Santha degree-relation routine from~\Cref{lem:degree-relation-routine}. Then, we describe an algorithm to find $s \in G$ as follows:
\begin{enumerate}
    \item Use the Cheung--Mosca algorithm~\cite{CheungMosca01} to compute an explicit cyclic decomposition of $G$, and fix a chain
    \begin{equation}\label{eq:abelian-chain}
        \hat G=A_0\supsetneq A_1\supsetneq\cdots\supsetneq A_k=\{0\}
    \end{equation}
    such that $A_{i-1}/A_i\cong\FF_{p_i}^{r_i}$ for all $i \in [k]$, where each $p_i$ is prime and $r_i\geq1$.
    \item Initialize $\chi$ to the unique character on $A_k=\{0\}$, that is, $\chi(0)=1$.
    \item Whenever a fresh binary Fourier sample is needed, apply~\Cref{lem:edcp-fourier-sampling} to a new coset state to obtain the binary Fourier sample $\left(a,\frac{1}{\sqrt2}\left(\ket0+\chi_s(a)\ket1 \right)\right)$, where $a$ is uniform in $\hat G$.
    \item For $i=k,k-1,\ldots,1$, with $\chi=\chi_s|_{A_i}$ at the start of the stage:
    \begin{enumerate}
        \item Use the explicit cyclic decomposition to find $t\in G$ such that $\chi_t|_{A_i}=\chi$. Apply~\Cref{lem:character-correction} to each fresh binary Fourier sample with $p=2$, $H=G$, $A=A_i$, $K=A_i^\perp$, and the extension $\chi_t$. For a sample with label $a$, the lemma's input function has $f(0)=0$ and $f(1)=a$: multiply the phase of $\ket1$ by $\chi_t(a)^{-1}$ and replace its label by $a+A_i$. The lemma's $r$ is $s-t\in A_i^\perp$, and the corrected samples have independent uniform labels in $\hat G/A_i$.
        \item For $j=1,\ldots,i-1$, turn binary samples with independent uniform labels in $A_{j-1}/A_i$ into binary samples with independent uniform labels in $A_j/A_i$ as follows:
        \begin{enumerate}
            \item Use~\Cref{lem:binary-to-p-ary-fourier-sample} with $p=p_j$, $H=A_i^\perp/A_{j-1}^\perp$, and secret $(s-t)+A_{j-1}^\perp$ to convert the current binary samples into $p_j$-ary Fourier samples. Here $\hat H\cong A_{j-1}/A_i$.
            \item Apply~\Cref{lem:modulus-reduction-correctness} with $p=p_j$, $r=r_j$, $H=A_i^\perp/A_{j-1}^\perp$, $A=A_j/A_i$, $K=A^\perp=A_j^\perp/A_{j-1}^\perp$, and secret $(s-t)+A_{j-1}^\perp$. Indeed, $\hat H/A\cong A_{j-1}/A_j\cong\FF_{p_j}^{r_j}$. Use the Ivanyos--Santha routine from~\Cref{lem:degree-relation-routine} for degrees $1,\ldots,p_j-1$; at degree one it takes $r_j+1$ coefficient vectors. The output $p_j$-ary samples have independent uniform label functions into $A_j/A_i$ and secret $(s-t)+A_j^\perp$ over $H/K\cong A_i^\perp/A_j^\perp$.
            \item Use~\Cref{lem:p-ary-to-binary-fourier-sample} with $p=p_j$, $H=A_i^\perp/A_j^\perp$, and secret $(s-t)+A_j^\perp$ to convert the outputs back to binary samples with independent uniform labels in $\hat H\cong A_j/A_i$.
        \end{enumerate}
        \item Use~\Cref{lem:binary-to-p-ary-fourier-sample} with $p=p_i$, $H=A_i^\perp/A_{i-1}^\perp$, and secret $(s-t)+A_{i-1}^\perp$ to convert independent binary samples with uniform labels in $\hat H\cong A_{i-1}/A_i$ into $p_i$-ary Fourier samples. When $i=1$, these binary samples come directly from the character correction in part~(a).
        \item Apply~\Cref{lem:terminal-recovery} with $p=p_i$, $r=r_i$, $H=A_i^\perp/A_{i-1}^\perp$, secret $(s-t)+A_{i-1}^\perp$, and coordinates $\hat H\cong A_{i-1}/A_i\cong\FF_{p_i}^{r_i}$. Its output determines the character of $A_{i-1}/A_i$ induced by $s-t$.
        \item Extend $\chi$ to $A_{i-1}$ by setting
        \[
            \chi(a)=\chi_t(a)\chi_{s-t}(a+A_i)
            \qquad\text{for every }a\in A_{i-1},
        \]
        where the second factor is the character recovered in part~(d).
    \end{enumerate}
    \item Identify $s$ from $\chi=\chi_s$ on $A_0=\hat G$ using the explicit cyclic coordinates, and output $s$.
\end{enumerate}
\end{algorithm}

\begin{theorem}
\label{thm:finite-abelian-hidden-shift}
Let $G$ be a finite abelian group. Suppose $\hat G$ has a chain with effective coordinates
\[
    \hat G=A_0\supsetneq A_1\supsetneq\cdots\supsetneq A_k=\{0\},
    \qquad A_{i-1}/A_i\cong\FF_{p_i}^{r_i}\quad(i\in[k]),
\]
where each $p_i$ is prime and $r_i\geq1$. For every $i$ and $2\leq d<p_i$, use the deterministic degree-$d$ relation routine with parameters $\ell(p_i,r_i,d)$ and $T(p_i,r_i,d)$ from~\Cref{lem:degree-relation-routine}. Then, with probability at least $\Omega(1)$, \Cref{algo:main} solves the generalized dihedral coset problem over $\dih(G)$ with at most $m$ samples and $T$ time, where
\begin{align*}
    m &= O \left(\prod_{i=1}^{k}
\left(8p_i(r_i+1)\prod_{d=2}^{p_i-1}
\ell(p_i,r_i,d)\right)\right), \text{ and } \\
T &= O\!\left(m\,\poly\!\left(
\log(m|G|),
\sum_{i=1}^{k}\left(p_i+
\sum_{d=2}^{p_i-1}\bigl(\ell(p_i,r_i,d)+T(p_i,r_i,d)\bigr)
\right)\right)\right).
\end{align*}
\end{theorem}

\begin{proof}
The explicit cyclic decomposition supplies the subgroup, quotient, and character operations needed in the algorithm. Each quotient $A_{i-1}/A_i$ has order $p_i^{r_i}\geq2$, so the chain has $k\leq\log_2|G|$ levels.
By~\Cref{lem:edcp-fourier-sampling}, Step 3 turns each fresh coset state into a binary Fourier sample with a uniform label in $\hat G$.

We claim that the algorithm maintains the following invariant: at the start of stage $i$ in the Step 4 loop, we have that $\chi = \chi_s|_{A_i}$. We prove this by induction. The base case $i=k$, when $\chi$ is initialized in Step 2, holds since the unique character on $A_k=\{0\}$ is $\chi_s|_{A_k}$.

By the induction hypothesis, suppose that at the start of stage $i$ in Step 4, we have that $\chi=\chi_s|_{A_i}$. The explicit cyclic decomposition isomorphism allows us to find some $t \in G$ such that $\chi_t|_{A_i}=\chi$. Hence $s-t\in A_i^\perp$.

\paragraph{Step 4(a). Correcting for the recovered parts of $s$.}
Applying~\Cref{lem:character-correction} in Step 4(a) with $p=2$ to fresh binary samples gives independent samples with secret $s-t$ and uniform labels in $\hat G/A_i$. 

\paragraph{Step 4(b). Modulus reduction from $A_0 := \hat G$ to $A_{i-1}$.}
At the start of pass $j$ in Step 4(b), the binary labels are independent and uniform in $A_{j-1}/A_i$. By~\Cref{lem:binary-to-p-ary-fourier-sample}, they give $p_j$-ary samples over $A_i^\perp/A_{j-1}^\perp$ with independent uniform label functions, with at most $4 p_j$ binary Fourier samples in expectation per output. 

In Step 4(b)(ii), for the current group $H=A_i^\perp/A_{j-1}^\perp$, the subgroup $A=A_j/A_i$ satisfies $\hat H/A\cong A_{j-1}/A_j\cong\FF_{p_j}^{r_j}$, and $K=A^\perp=A_j^\perp/A_{j-1}^\perp$. The Ivanyos--Santha routine from~\Cref{lem:degree-relation-routine} finds a nonzero degree-one relation among $r_j+1$ coefficient vectors. Thus~\Cref{lem:modulus-reduction-correctness} gives a $p_j$-ary sample over $H/K\cong A_i^\perp/A_j^\perp$ with secret $(s-t)+A_j^\perp$ and a uniform label function into $A_j/A_i$. It uses at most
\[
(r_j+1)\prod_{d=2}^{p_j-1}\ell(p_j,r_j,d)
\]
input $p_j$-ary samples. Calls on disjoint fresh inputs give independent outputs. By~\Cref{lem:p-ary-to-binary-fourier-sample}, a successful conversion of an output of part~(ii) gives a binary sample with a uniform label in $A_j/A_i$. The conversion succeeds with probability at least $2/3$, so it uses fewer than two $p_j$-ary samples in expectation. Both conversion success probabilities are independent of the labels, so retrying with fresh samples preserves the uniform independent output labels. Combining steps 4(b)(i)--(iii), one binary output of pass $j$ costs fewer than $8p_j(r_j+1)\prod_{d=2}^{p_j-1}\ell(p_j,r_j,d)$ binary inputs from the preceding pass in expectation.

\paragraph{Step 4(c).} After pass $i-1$, the binary labels are independent and uniform in $A_{i-1}/A_i$; when $i=1$, this follows directly from part~(a). By~\Cref{lem:binary-to-p-ary-fourier-sample}, fewer than $4p_i$ such binary samples in expectation give one $p_i$-ary sample with a uniform label function.

\noindent\textbf{Step 4(d).} The dual of $A_i^\perp/A_{i-1}^\perp$ is $A_{i-1}/A_i\cong\FF_{p_i}^{r_i}$. Thus~\Cref{lem:terminal-recovery} recovers the character induced by $s-t$ on this quotient with probability $1$. It uses at most
\[
(r_i+2)\prod_{d=2}^{p_i-1}\ell(p_i,r_i,d)
\]
$p_i$-ary samples in expectation. With Step 4(c), the expected number of binary samples needed after the inner loop is less than $4p_i(r_i+2)\prod_{d=2}^{p_i-1}\ell(p_i,r_i,d)$.

\noindent\textbf{Step 4(e).} Since $s-t\in A_i^\perp$, its character is well defined on $A_{i-1}/A_i$. For every $a\in A_{i-1}$, the update in the algorithm therefore gives
\[
\chi(a)=\chi_t(a)\chi_{s-t}(a+A_i)=\chi_t(a)\chi_{s-t}(a)=\chi_s(a).
\]
Hence $\chi=\chi_s|_{A_{i-1}}$, as required for the next stage.

Therefore, after stage $i=1$, we know $\chi_s$ on $A_0=\hat G$. The explicit cyclic decomposition and \Cref{thm:pontryagin} allow us to find the unique $s\in G$ with this character.

\paragraph{Time and sample complexity.}
The expected sample costs multiply because each pass combines samples produced by the preceding pass, and summing over recovery stages changes the full product by only a constant factor, giving the stated bound $m$ in expectation. The calls to \cite{IvanyosSantha16} and polynomial work per sample give the corresponding expected running time $T$. Aborting runs that exceed sufficiently large constant multiples of these expectations gives the stated time and sample bounds with constant success probability by Markov's inequality.
\end{proof}

\begin{corollary}[Finite abelian groups of smooth exponent]
\label{cor:abelian-smooth-exponent}
Let $G$ be a finite abelian group whose exponent is $B$-smooth, where $B\geq2$. Suppose $\hat G$ has a chain with effective coordinates $\hat G=A_0\supsetneq\cdots\supsetneq A_k=\{0\}$ and $A_{i-1}/A_i\cong\FF_{p_i}^{r_i}$ for every $i\in[k]$, where each $p_i$ is prime and $r_i\geq1$. Then, for every $0<\varepsilon<1$, the generalized dihedral coset problem over $\dih(G)$ can be solved with probability at least $1-\varepsilon$. Both its sample complexity and running time are at most
\[
\poly\!\left(B,\log|G|,\log\frac{2}{\varepsilon}\right)
B^{O(kB^{2+o(1)})}
\prod_{i=1}^k(1+r_i)^{O(B^{1+o(1)})}.
\]
\end{corollary}

\begin{proof}
Fix a quotient $\FF_p^r$ and a degree $d$, and put $g=\gcd(d,p-1)$ as in~\Cref{lem:degree-relation-routine}. Since $g<p$ and $\lceil\log_2(g+1)\rceil=O(\log p)$, its number of input vectors satisfies
\[
\ell(p,r,d)
\leq(r+1)^{O(g\log p)}p^{O(g^2\log p)}.
\]
This also holds when $g=1$, for which $\ell(p,r,d)=r+1$.

For each divisor $g$ of $p-1$, at most $(p-1)/g$ degrees have $\gcd(d,p-1)=g$, since each such degree is a multiple of $g$. Grouping the factors by this gcd gives
\begin{align*}
\prod_{d=2}^{p-1}\ell(p,r,d)
&\leq(r+1)^{O\left(p\log p\,\#\{g\geq1:g\mid p-1\}\right)}
p^{O\left(p\log p\sum_{g\mid p-1}g\right)}\\
&\leq(r+1)^{O(p^{1+o(1)})}p^{O(p^{2+o(1)})}.
\end{align*}
For the second inequality, \Cref{lem:divisor-bound} gives $\#\{g\geq1:g\mid p-1\}=p^{o(1)}$, and each divisor is less than $p$, so $\sum_{g\mid p-1}g\leq p\,\#\{g\geq1:g\mid p-1\}$. The factor $\log p$ is absorbed into $p^{o(1)}$. When $p=2$, the product is empty and the same bound holds.

Multiplying by $8p(r+1)$ preserves the last bound, so~\Cref{thm:finite-abelian-hidden-shift} bounds the sample complexity by the product of these factors over the chain. Every $p_i\leq B$, giving the stated separate powers of $B$ and $1+r_i$. The time bound has the same form: $T(p,r,d)$ is polynomial in $\ell(p,r,d)$ and $p$, and the theorem's additional polynomial work only changes the absolute constants in these exponents.

Repeating the capped algorithm $O(\log(2/\varepsilon))$ times and allocating total approximation error at most $\varepsilon/2$ gives the stated success probability and costs.

\end{proof}

The vector DCP guarantee follows by taking the subgroup chain that removes one prime factor of $q$ at a time.

\begin{corollary}[Vector DCP over smooth moduli]
\label{thm:edcp-correctness}
Let $n\geq1$, let $q>1$ be $B$-smooth, and let $0<\varepsilon<1$. A quantum algorithm recovers every $\vecs\in\ZZ_q^n$ from independent vector DCP samples with probability at least $1-\varepsilon$. Both its sample complexity and running time are at most
\[
n^{O(B^{1+o(1)}\log q)}B^{O(B^{2+o(1)}\log q)}
\poly\!\left(\log\frac{2}{\varepsilon}\right).
\]
\end{corollary}

\begin{proof}
Write $q=\prod_{i=1}^k p_i$, with repetition, and take $A_0=\widehat{\ZZ_q^n}\cong\ZZ_q^n$ and $A_i=(p_1\cdots p_i)A_0$. Then $A_{i-1}/A_i\cong\FF_{p_i}^n$, the coordinates are given by modular arithmetic, and $k\leq\log_2q$. Apply \Cref{cor:abelian-smooth-exponent} with $r_i=n$ and $|G|=q^n$. Its product over the chain is at most $n^{O(B^{1+o(1)}\log q)}B^{O(B^{2+o(1)}\log q)}$. The polynomial overhead is absorbed into this bound, apart from the stated dependence on $\varepsilon$.

\end{proof}

\bibliographystyle{alpha}
\bibliography{refs}

@article{ImranIvanyos23,
  author        = {Muhammad Imran and G{\'a}bor Ivanyos},
  title         = {Zero Sum Subsequences and Hidden Subgroups},
  journal       = {arXiv preprint arXiv:2304.08376},
  year          = {2023}
}

@article{IvanyosSantha16,
  author        = {G{\'a}bor Ivanyos and Mikl{\'o}s Santha},
  title         = {Solving Systems of Diagonal Polynomial Equations over Finite Fields},
  journal       = {Theoretical Computer Science},
  volume        = {657},
  pages         = {73--85},
  year          = {2017},
  doi           = {10.1016/j.tcs.2016.04.045}
}

@article{KothariODonnellWu26,
  author        = {Robin Kothari and Ryan O'Donnell and Kewen Wu},
  title         = {No Exponential Quantum Speedup for {$\mathsf{SIS}^{\infty}$}
                   Anymore},
  journal       = {arXiv preprint arXiv:2510.07515},
  year          = {2026}
}

@article{BaiEtAl25,
  author        = {Shi Bai and Hansraj Jangir and Elena Kirshanova and
                   Tran Ngo and William Youmans},
  title         = {A Quasi-polynomial Time Algorithm for the Extrapolated
                   Dihedral Coset Problem over Power-of-Two Moduli},
  journal       = {IACR Cryptology ePrint Archive, Paper 2025/1046},
  url           = {https://eprint.iacr.org/2025/1046},
  year          = {2025}
}

@inproceedings{brakerski2018learning,
  title={Learning with errors and extrapolated dihedral cosets},
  author={Brakerski, Zvika and Kirshanova, Elena and Stehl{\'e}, Damien and Wen, Weiqiang},
  booktitle={IACR international workshop on public key cryptography},
  pages={702--727},
  year={2018},
  organization={Springer}
}

@inproceedings{ChenLiuZhandry22,
  author    = {Yilei Chen and Qipeng Liu and Mark Zhandry},
  title     = {Quantum Algorithms for Variants of Average-Case Lattice Problems via Filtering},
  booktitle = {Advances in Cryptology -- EUROCRYPT 2022, Part III},
  series    = {Lecture Notes in Computer Science},
  volume    = {13277},
  pages     = {372--401},
  publisher = {Springer},
  year      = {2022},
  doi       = {10.1007/978-3-031-07082-2_14}
}

@article{Regev04,
  author  = {Oded Regev},
  title   = {Quantum Computation and Lattice Problems},
  journal = {SIAM Journal on Computing},
  volume  = {33},
  number  = {3},
  pages   = {738--760},
  year    = {2004},
  doi     = {10.1137/S0097539703440678}
}

@inproceedings{Kuperberg13,
  author    = {Greg Kuperberg},
  title     = {Another Subexponential-Time Quantum Algorithm for the
               Dihedral Hidden Subgroup Problem},
  booktitle = {8th Conference on the Theory of Quantum Computation,
               Communication and Cryptography (TQC 2013)},
  series    = {LIPIcs},
  volume    = {22},
  pages     = {20--34},
  year      = {2013}
}

@article{ChildsJaoSoukharev14,
  author  = {Andrew M. Childs and David Jao and Vladimir Soukharev},
  title   = {Constructing Elliptic Curve Isogenies in Quantum
             Subexponential Time},
  journal = {Journal of Mathematical Cryptology},
  volume  = {8},
  number  = {1},
  pages   = {1--29},
  year    = {2014},
  doi     = {10.1515/jmc-2012-0016}
}

@inproceedings{BonnetainSchrottenloher20,
  author    = {Xavier Bonnetain and Andr{\'e} Schrottenloher},
  title     = {Quantum Security Analysis of {CSIDH}},
  booktitle = {Advances in Cryptology---EUROCRYPT 2020, Part II},
  series    = {Lecture Notes in Computer Science},
  volume    = {12106},
  pages     = {493--522},
  publisher = {Springer},
  year      = {2020},
  doi       = {10.1007/978-3-030-45724-2_17}
}

@inproceedings{Peikert20,
  author    = {Chris Peikert},
  title     = {He Gives {C}-Sieves on the {CSIDH}},
  booktitle = {Advances in Cryptology---EUROCRYPT 2020, Part II},
  series    = {Lecture Notes in Computer Science},
  volume    = {12106},
  pages     = {463--492},
  publisher = {Springer},
  year      = {2020}
}

@article{EttingerHoyerKnill04,
  author  = {Mark Ettinger and Peter H{\o}yer and Emanuel Knill},
  title   = {The Quantum Query Complexity of the Hidden Subgroup Problem
             is Polynomial},
  journal = {Information Processing Letters},
  volume  = {91},
  number  = {1},
  pages   = {43--48},
  year    = {2004}
}

@article{Kuperberg05,
  author  = {Greg Kuperberg},
  title   = {A Subexponential-Time Quantum Algorithm for the Dihedral
             Hidden Subgroup Problem},
  journal = {SIAM Journal on Computing},
  volume  = {35},
  number  = {1},
  pages   = {170--188},
  year    = {2005},
  doi     = {10.1137/S0097539703436345}
}

@article{CheungMosca01,
  author  = {Kevin K. H. Cheung and Michele Mosca},
  title   = {Decomposing Finite Abelian Groups},
  journal = {Quantum Information and Computation},
  volume  = {1},
  number  = {3},
  pages   = {26--32},
  year    = {2001},
  eprint  = {cs/0101004}
}

@article{Regev04DHSP,
  author  = {Oded Regev},
  title   = {A Subexponential Time Algorithm for the Dihedral Hidden
             Subgroup Problem with Polynomial Space},
  journal = {arXiv preprint quant-ph/0406151},
  year    = {2004}
}

@article{FriedlEtAl14,
  author  = {Katalin Friedl and G{\'a}bor Ivanyos and Fr{\'e}d{\'e}ric
             Magniez and Miklos Santha and Pranab Sen},
  title   = {Hidden Translation and Translating Coset in Quantum Computing},
  journal = {SIAM Journal on Computing},
  volume  = {43},
  number  = {1},
  pages   = {1--24},
  year    = {2014},
  doi     = {10.1137/130907203}
}

@article{Ivanyos08,
  author  = {G{\'a}bor Ivanyos},
  title   = {On Solving Systems of Random Linear Disequations},
  journal = {Quantum Information and Computation},
  volume  = {8},
  number  = {6--7},
  pages   = {579--594},
  year    = {2008}
}

@inproceedings{BonnetainNayaPlasencia18,
  author    = {Xavier Bonnetain and Mar{\'i}a Naya-Plasencia},
  title     = {Hidden Shift Quantum Cryptanalysis and Implications},
  booktitle = {Advances in Cryptology---ASIACRYPT 2018, Part I},
  series    = {Lecture Notes in Computer Science},
  volume    = {11272},
  pages     = {560--592},
  publisher = {Springer},
  year      = {2018},
  doi       = {10.1007/978-3-030-03326-2_19}
}

@article{Csaji21,
  author  = {Gergely Cs{\'a}ji},
  title   = {A New Quantum Algorithm for the Hidden Shift Problem in
             {$\mathbb{Z}_{2^t}^n$}},
  journal = {arXiv preprint arXiv:2102.04171},
  year    = {2021}
}

@inproceedings{CastryckEtAl21,
  author    = {Wouter Castryck and Ann Dooms and Carlo Emerencia and
               Alexander Lemmens},
  title     = {A Fusion Algorithm for Solving the Hidden Shift Problem in
               Finite Abelian Groups},
  booktitle = {Post-Quantum Cryptography---PQCrypto 2021},
  series    = {Lecture Notes in Computer Science},
  volume    = {12841},
  pages     = {133--153},
  publisher = {Springer},
  year      = {2021},
  doi       = {10.1007/978-3-030-81293-5_8}
}

@article{simon1997power,
  title={On the power of quantum computation},
  author={Simon, Daniel R},
  journal={SIAM journal on computing},
  volume={26},
  number={5},
  pages={1474--1483},
  year={1997},
  publisher={SIAM}
}

@article{boucher2026cyclotomic,
  title={Cyclotomic Cosets: Hidden Subgroup and Quantum Sieving Algorithm for Prime-Power Moduli},
  author={Boucher, Mathias and Fouque, Pierre-Alain and Shen, Yixin},
  journal={arXiv preprint arXiv:2609.34996},
  year={2026}
}

\end{document}